\documentclass[12pt]{article}
\pdfoutput=1
\usepackage{epsf}
\usepackage{amsmath}
\usepackage{amsfonts}
\usepackage{amssymb}
\usepackage{graphicx}
\usepackage{color}
\usepackage{psfrag}
\usepackage{cite}
\usepackage{hyperref}
\usepackage{tikz}

\usepackage{ifpdf}

\newcommand{\bmat}{\left(\begin{array}}
\newcommand{\emat}{\end{array}\right)}

\def\yzero{\smash{\hbox{$y\kern-4pt\raise1pt\hbox{${}^\circ$}$}}}

\def\beq{\begin{equation}}
\def\eeq{\end{equation}}
\def\beqa{\begin{eqnarray}}
\def\eeqa{\end{eqnarray}}

\def\-{\hphantom{-}}

\def\s2{\frac{1}{\sqrt2}}

\def\beq{\begin{equation}}
\def\eeq{\end{equation}}
\def\beqa{\begin{eqnarray}}
\def\eeqa{\end{eqnarray}}

\def\Tr{{\rm Tr \,}}

\def\IF{\relax{\rm I\kern-.18em F}}
\def\II{\relax{\rm I\kern-.18em I}}

\def\Dsl{\,\raise.15ex\hbox{/}\mkern-13.5mu D} 

\def\IS{{\bf S}}
\def\IR{{\bf R}}
\def\IZ{{\bf Z}}

\def\CF{{\cal F}}

\newcommand{\eq}[1]{(\ref{#1})}

\catcode`\@=11   
\newdimen\@rotdimen
\newbox\@rotbox  

\def\@vspec#1{\special{ps:#1}}
\def\@rotstart#1{\@vspec{gsave currentpoint currentpoint translate
   #1 neg exch neg exch translate}}
\def\@rotfinish{\@vspec{currentpoint grestore moveto}}
\def\@rotr#1{\@rotdimen=\ht#1\advance\@rotdimen by\dp#1%
   \hbox to\@rotdimen{\hskip\ht#1\vbox to\wd#1{\@rotstart{90 rotate}%
   \box#1\vss}\hss}\@rotfinish}
\def\@rotl#1{\@rotdimen=\ht#1\advance\@rotdimen by\dp#1%
   \hbox to\@rotdimen{\vbox to\wd#1{\vskip\wd#1\@rotstart{270 rotate}%
   \box#1\vss}\hss}\@rotfinish}%
\def\@rotu#1{\@rotdimen=\ht#1\advance\@rotdimen by\dp#1%
   \hbox to\wd#1{\hskip\wd#1\vbox to\@rotdimen{\vskip\@rotdimen
   \@rotstart{-1 dup scale}\box#1\vss}\hss}\@rotfinish}%
\def\@rotf#1{\hbox to\wd#1{\hskip\wd#1\@rotstart{-1 1 scale}%
   \box#1\hss}\@rotfinish}%
\def\rotate{\@ifnextchar[{\@rotate}{\@rotate[l]}}
\def\@rotate[#1]#2{\setbox\@rotbox=\hbox{#2}\@nameuse{@rot#1}\@rotbox}

\catcode`\@=12

\topmargin
-1.5cm
\textwidth
15.5cm
\textheight
23.5cm
\oddsidemargin
0.7cm
\evensidemargin
1.2cm

\begin{document}

\makeatletter
\@addtoreset{equation}{section}
\makeatother
\renewcommand{\theequation}{\thesection.\arabic{equation}}
\pagestyle{empty}
\rightline{ IFT-UAM/CSIC-15-129}
\rightline{ MPP-2015-281}
\vspace{1.2cm}
\begin{center}
\LARGE{\bf Relaxion Monodromy \\
and the Weak Gravity Conjecture \\[12mm] }
\large{L. E. Ib\'a\~nez$^{1,2}$, M. Montero$^{1,2}$,  A. M. Uranga$^2$, I. Valenzuela$^{3,4}$\\[4mm]}
\footnotesize{${}^{1}$ Departamento de F\'{\i}sica Te\'orica, Facultad de Ciencias\\[-0.3em] 
Universidad Aut\'onoma de Madrid, 28049 Madrid, Spain\\
${}^2$ Instituto de F\'{\i}sica Te\'orica IFT-UAM/CSIC,\\[-0.3em] 
C/ Nicol\'as Cabrera 13-15, 
Campus de Cantoblanco, 28049 Madrid, Spain} \\ 
${}^3$ Max-Planck-Institut fur Physik,
Fohringer Ring 6, 80805 Munich, Germany\\
${}^4$Institute for Theoretical Physics and
Center for Extreme Matter and Emergent Phenomena,\\
Utrecht University, Leuvenlaan 4, 3584 CE Utrecht, The Netherlands\\

\vspace*{5mm}

\small{\bf Abstract} \\
\end{center}
\begin{center}
\begin{minipage}[h]{17.0cm}
The recently proposed relaxion models require extremely large trans-Planckian axion excursions as well as a
potential explicitly violating the axion shift symmetry. The latter property is however inconsistent with the axion periodicity, which corresponds to a gauged discrete shift symmetry. A way to make things consistent is to use monodromy, i.e. both the axion and the potential parameters transform under the discrete shift symmetry. The structure is better described in terms of  a 3-form field $C_{\mu \nu \rho}$ coupling to the SM Higgs through its field strength $F_4$.
The 4-form also couples linearly to the relaxion, in the Kaloper-Sorbo fashion.
The extremely small  relaxion-Higgs coupling arises in a see-saw fashion as
$g\simeq F_4/f$, with $f$ being the axion decay constant.
 We discuss constraints on this type of constructions from membrane nucleation and the Weak Gravity Conjecture. 
 The latter requires the existence of membranes, whose too fast nucleation could in principle drive the theory out of control, unless the cut-off scale is lowered. This allows to constrain relaxion models on purely theoretical grounds. 
We also discuss possible avenues to embed this structure into string theory.

\end{minipage}
\end{center}
\newpage
\setcounter{page}{1}
\pagestyle{plain}
\renewcommand{\thefootnote}{\arabic{footnote}}
\setcounter{footnote}{0}

\tableofcontents 

\vspace*{1cm}
\section{Introduction}
Recently \cite{Graham:2015cka} (see also \cite{Espinosa:2015eda,Hardy:2015laa,Patil:2015oxa,Jaeckel:2015txa,Gupta:2015uea,Batell:2015fma,Marzola:2015dia,Choi:2015fiu,Kaplan:2015fuy,DiChiara:2015euo} 
 for later versions) proposed a new mechanism to solve the EW hierarchy problem under the name of \emph{cosmological relaxation}. Its main appeal is that it does not require the presence of new physics near the EW scale, while  providing at the same time a natural dynamical mechanism to keep the Higgs hierarchically lighter than the cutoff of the theory. Basically the proposal is to extend the SM Higgs scalar potential by including a coupling to an axionic field, leading to a potential
\beq
V=V(g\phi)+(-M^2+g\phi)|h|^2+\Lambda^4\cos\left(\frac{\phi}{f}\right).
\label{relaxation}
\eeq
Here $V(g\phi)=gM^2\phi+g^2\phi^2+\dots$, and $\Lambda=\Lambda_0+\Lambda(h)$ depends on the vev of the Higgs field $h$. Also, $f$ is the usual axion decay constant, and  $M^2$ is a cutoff coming from SM loop effects. 

In the minimal version the field $\phi$ is the QCD axion, the cosine potential arises from the usual $SU(3)$ instanton effects breaking the Peccei-Quinn symmetry, and $\Lambda_{QCD}\sim \Lambda(h=v)$ depends on the Higgs vev through quark masses. We will however consider 
also more general axion-like particles to play the relaxion role.

During inflation, the non-perturbative effects are negligible, hence the dynamics of $\phi$ is controlled by $V(g\phi)$; the axion $\phi$ starts out at a large positive value and slow rolls down its potential, thus scanning values for the Higgs mass. When crossing $m_h\sim 0$, namely at $\phi\sim M^2/g$, the Higgs develops a vev, triggering electroweak symmetry breaking, and the barrier of the cosine potential increases stabilizing the axion shortly after $m_h\sim 0$. Hence the Higgs mass is dynamically set to a value much lower than the cutoff $M$. In order for the instanton term to stop the rolling of $\phi$, the barrier $\Lambda(h)$ evaluated at $h=v$ should be comparable to the slope of the axion potential. Parametrizing $\Lambda(h)^4=ch^2$, this requires
\beq
gM^ 2\sim \frac{\Lambda^4(h=v)}{f}\quad \longrightarrow\quad  g\sim \frac{cv^2}{fM^2}\label{axcop}
\eeq
For the relaxion being the QCD axion, $c=f^2_\pi y^2_u$, and $f>10^9$ GeV according to astrophysical bounds, leading to a very small coupling 
\beq
g\sim 10^{-16}\frac{m_{\rm EW}^2}{M^2}\text{ GeV}.
\eeq
Therefore a big hierarchy between the cutoff $M$ and the EW scale is translated into a very small coupling $g$. In fact, the smallness of this parameter is common in all the versions of the cosmological relaxation mechanism, with $g\sim 10^{-34}$ GeV being a typical value.

This mechanism to generate a hierarchically small Higgs mass is argued to be technically natural, since the smallness of $m_h$ comes from the smallness of the parameters $g$ and $\Lambda$, which are associated to symmetry breakings. Indeed, the parameter $g$ is  the only source of breaking of the global shift symmetry of the axion, and therefore its smallness is expected to be technically natural.

There are two important questions unanswered by the above description:

$\bullet$ The smallness of $g$ implies a field excursion $\Delta\phi$ during inflation much larger than the UV cutoff of the theory. This possibly endangers the stability of the potential against higher dimensional operators, a familiar issue in large field inflation models (see \cite{Baumann:2014nda} for a review). One may argue that this problem is solved  by appealing to  the continuous perturbative axion shift symmetry. However, given the general belief that quantum gravity violates all global symmetries, this mechanism seems unrealizable in actual embedding of this effective theory in UV completions including quantum gravity.

$\bullet$ On the other hand, for $\phi$ to describe an axion, it should have a discrete periodic identification under $\phi\rightarrow \phi+2\pi f$. As emphasized in \cite{Gupta:2015uea}, this should correspond to a  gauge symmetry in a consistent theory of quantum gravity \footnote{One could try to drop the requirement that $\phi$ is an axion, by declaring it to take values in $\IR$ rather than in $\IS^1$. This however makes the PQ-symmetry group a ( nonlinearly realized) $\IR$, instead of the usual $\IS^1=U(1)$. Noncompact symmetries are again notoriously in conflict with quantum gravity \cite{Banks:2010zn}.}. This is however not respected by the coupling to the Higgs field in \eq{relaxation}, implying that $\phi$ can not be a pseudo-Nambu-Goldstone boson (pNGB).

\medskip

It is interesting that these two questions have already been addressed in the context of axion monodromy inflation models
\cite{Silverstein:2008sg,McAllister:2008hb,Kaloper:2008fb,Kaloper:2014zba,Marchesano:2014mla,McAllister:2014mpa} (see also \cite{ Berg:2009tg, Ibanez:2014zsa,Palti:2014kza,Blumenhagen:2014gta,Ibanez:2014kia,Hebecker:2014eua,Arends:2014qca,Franco:2014hsa,Blumenhagen:2014nba,Hayashi:2014aua,Hebecker:2014kva,Ibanez:2014swa,Garcia-Etxebarria:2014wla,Blumenhagen:2015kja,Retolaza:2015sta,Escobar:2015fda,Bielleman:2015lka,Blumenhagen:2015xpa}),
and this motivates the suggestion to UV-complete relaxion models in this framework. In this paper we take several important steps towards fleshing out this proposal, and analyzing the problems of embedding it into a consistent quantum theory of gravity like string theory. Among other things, we will study the constraints that the Weak Gravity Conjecture \cite{ArkaniHamed:2006dz}
implies in the viability of the models. 
 Inflation generically leads to too fast nucleation of the membranes required by the WGC, which could drive  the theory beyond control if the 
 cut-off scale is not low enough.
We also consider possible string theory embeddings of the relaxion monodromy potentials.

The paper is organized as follows. In section \ref{sec:axion-monodromy} we review axion monodromy, i.e. periodic scalars with multibranched potentials. In section \ref{sec:toy-model} we describe a minimal relaxion model where the potential, including the Higgs coupling, arises from the multibranched monodromic structure. section \ref{sec:seesaw} highlights an interesting parametric dependence of scales, which can naturally accommodate the exceedingly small couplings of relaxion models; section \ref{sec:coupling-multibranched} introduces a simple  multibranched relaxion model. In section \ref{WGCmon}  we discuss nucleation of membranes bounding bubbles of vacua corresponding to different branches of the axion potential,  and exploit the possible role of the Weak Gravity Conjecture in the evaluation of the 
transition rates.  This discussion is applied to the case of relaxion models in section \ref{sec:killrel}. In section \ref{sec:string-relaxion} we discuss possible 
embeddings of the relaxion structure into a string theory setting. We finally leave section \ref{sec:conclusions} for general comments and conclusions.
We complete the main text with four appendices.
Appendix \ref{sec:Kaloper-Sorbo} reviews axion monodromy in terms of a dual 3-form with a 2-form gauge symmetry. In appendix 
 \ref{sec:bubbles} we derive the tunnelling probability formulae discussed in the text. Appendix \ref{WGC3} discusses the Weak Gravity Conjecture 
 as applied to $(d-1)$-form gauge fields. Finally appendix \ref{sec:DBI} shows some details of the axion potential derived from the 
 DBI action, as used in section \ref{sec:string-relaxion}.

\section{Axion monodromy}
\label{sec:axion-monodromy}

In \cite{Gupta:2015uea} the authors argue that the discrete shift symmetry of $\phi$ has to be necessarily gauged in a consistent quantum theory of gravity and therefore can not be broken by any term in the action. This implies that if $\phi$ is an axion or a pseudo-Nambu-Goldostone boson (pNGB), the coupling $g$ is not naturally small but indeed theoretically inconsistent. However, the authors in \cite{Gupta:2015uea} 
miss the possibility that $\phi$ is not a pNGB but an \emph{axion with multi-branched potential}, so that the theory is consistent with a mass term and interactions for the axions while preserving an underlying discrete shift symmetry (see e.g. \cite{Silverstein:2008sg,McAllister:2008hb,Kaloper:2008fb,McAllister:2014mpa,Marchesano:2014mla,Kaloper:2014zba}, also  \cite{Berg:2009tg,Ibanez:2014zsa,Palti:2014kza,Blumenhagen:2014gta,Ibanez:2014kia, Hebecker:2014eua,Arends:2014qca,Franco:2014hsa,Blumenhagen:2014nba,Hayashi:2014aua,Hebecker:2014kva,Ibanez:2014swa,Garcia-Etxebarria:2014wla,Blumenhagen:2015kja,Retolaza:2015sta,Escobar:2015fda,Bielleman:2015lka} for applications to inflationary potentials). Our present work is the first concrete proposal to implement a monodromy realization of relaxion models, and explore its implications.

In this section we review axion monodromy models, explaining the mechanism by which periodic scalars get multi-branched potentials from the introduction of a coupling to a 3-form field. It also serves to fix notation and conventions.

As described in  \cite{Kaloper:2008fb} (see \cite{Dvali:2005an} for related ideas in a different context), an efficient way to describe the introduction of potential terms for axionic scalars is to couple them to a 3-form gauge field. Consider for instance the simplest case, which eventually describes a massive axion. It corresponds to the lagrangian
\beq
L=-\frac12(\partial_{\mu} \phi)^2-\frac12|F_4|^2+g\phi F_4,
\label{KSpot}
\eeq
where 
$F_4=dC_3$ the field strength of the 3-form. Since the 3-form field has no propagating degrees of freedom in 4d, we can integrate it out via its equation of motion
\beq
*F_4=f_0+g\phi,
\eeq
leading to an induced scalar potential for the axion
\beq
V_{KS}=\frac12(f_0+g\phi)^2.
\label{V0}
\eeq
Notice that even if the 3-form in four dimensions does not have propagating degrees of freedom, it can still yield a non-vanishing field strength giving a positive contribution $f_0$ to the vacuum energy.
The discrete identification of the scalar is a gauge symmetry which involves a change in $f_0$, as follows
\beq
\phi\rightarrow \phi+2\pi f\quad ,\quad f_0\rightarrow f_0-2\pi gf
\label{transf}
\eeq

At the quantum level\footnote{At the classical level, $f_0$ can take an arbitrary constant value implying that the continuous shift of the axion is also a symmetry of the action. However, as emphasized in \cite{Bousso:2000xa} the actual value of the 4-form field strength in four dimensions (and not only its shift when crossing a membrane) satisfies Dirac quantization. When embedding the model in string theory, this quantized value indeed corresponds to the integer flux of the magnetic dual in higher dimensions \cite{Bousso:2000xa,Bielleman:2015ina}}, the vacuum value of the 4-form flux $f_0$ is quantized in units of membrane charge (we will come back to these membranes in section \ref{WGCmon})
\beq
f_0=n\Lambda^2_k\quad , \quad n\in\IZ
\label{f0}
\eeq
Hence we have the following consistency condition \cite{Kaloper:2011jz}
\beq
2\pi f g = k \Lambda^2_k\quad ,\quad k\in {\IZ}
\label{qmf}
\eeq
which will be important when discussing explicit relaxion monodromy models.

\begin{figure}[t]
\begin{center}
\includegraphics[width=10cm]{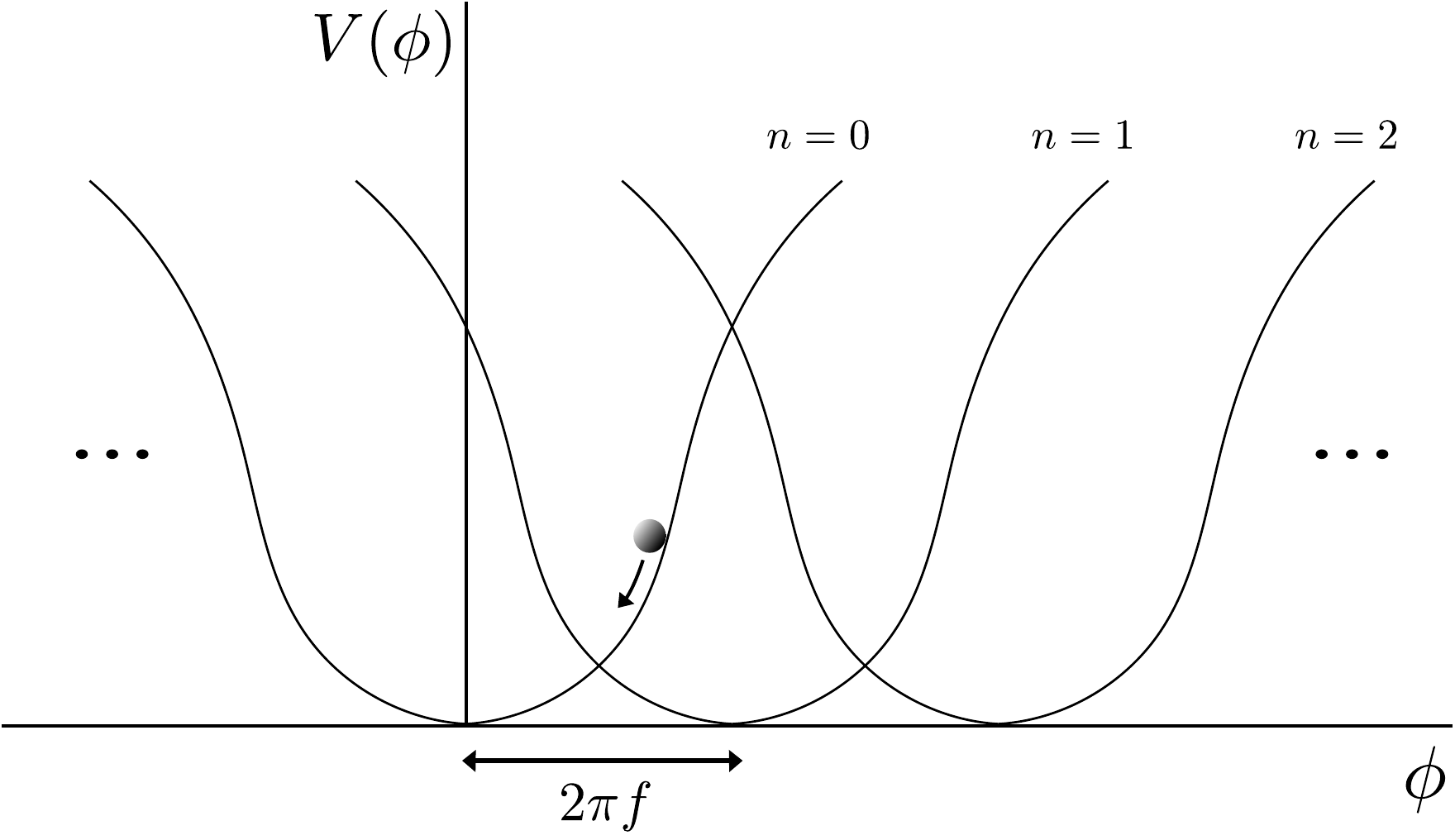}
\end{center}
\caption{Multi-branched structure of a typical axion monodromy model.}
\label{ramas}
\end{figure}

This structure underlies the axion monodromy inflationary models (see e.g. \cite{Silverstein:2008sg,McAllister:2008hb,Kaloper:2008fb,Kaloper:2014zba,Marchesano:2014mla,McAllister:2014mpa}), in which the scalar potential is multivalued with a multibranched structure dictated by the above discrete shift symmetry, akin to the ``repeated zone scheme'' familiar from solid-state physics \cite{Kaloper:2011jz,Ashcroft} (see Figure \ref{ramas} for a qualitative picture). Each branch is labelled by the value of $f_0$. Once a specific branch is chosen, one can go up in the potential away from the minimum and travel a distance $\Delta \phi$  larger than the fundamental periodicity $f$. This is specially useful for large field inflationary models in which one needs a trans-Planckian field excursion for the inflation even if all the scales of the theory remain sub-Planckian. The relation between F-term axion monodromy and a \emph{Kaloper-Sorbo} (KS) potential like \eqref{KSpot} was explicitly shown in \cite{Marchesano:2014mla}, and further generalised  in \cite{Bielleman:2015ina} for any axion of a given string compatification.

The above system can also be generalised to multiple axions
\beq
L=K_{ij}d \phi^i\wedge *d\phi^j-F_4\wedge *F_4+C_3\wedge j
\eeq
where $j_\mu=\partial_{\mu}\theta(\phi^i)$ is an external current satisfying $dj=0$ and $\theta(\phi_i)$ can be a polynomial function mixing the axions. This leads to more generic potentials than \eqref{V0} containing also quartic or higher couplings between the axions. The system is then invariant under a set of discrete transformations in the axions combined with some integer shifts on the parameters appearing in $\theta(\phi_i)$ (which in string theory correspond to the internal fluxes of the compactification). The relation \eqref{qmf} then becomes
\beq
\delta \theta(\phi)= k \Lambda_k^2\quad ,\quad k\in {\IZ}
\eeq
where $\delta \theta(\phi)$ is the integer shift that has to be reabsorbed by each 4-form background. This is indeed the situation that naturally arises in string theory flux compactifications \cite{Bielleman:2015ina}, in which the full axionic dependence of the scalar potential can be written in terms of 4d couplings of the 4-forms.

\smallskip

The above system of a single axion admits an alternative description in which the scalar is dualized into a 2-form. The 3-form then gets massive by ``eating up" the 2-form in a gauge invariant consistent way \cite{Marchesano:2014mla}. We review this (standard) description in appendix \ref{sec:Kaloper-Sorbo}. The description in terms of 2-forms is valid for the simple monodromic models studied in this paper, but has not appeared in the literature for more general monodromic models such as those in \cite{Bielleman:2015ina}.

The appealing feature of this mechanism is that the gauge invariance of the 3-form protects the potential from Planck-suppressed operators. More concretely, consider  higher-dimensional operators which appear as powers of the gauge invariant field strength $F_4/M^2$, with $M$ the cut-off scale (or the Planck mass in
monodromy inflation). After integrating out the 3-form  the corrections to the scalar potential will also appear as powers of the leading order potential itself, namely $\delta V\sim  \sum_n V_0(V_0/M^4)^n$, so they will be subleading as long as the potential remains below the cut-off scale, even if the field takes large values. Therefore it is a very efficient mechanism to keep a scalar field  light, in a way consistent with interactions,  and without adding new degrees of freedom or new physics at the EW scale.

This is another motivation to  construct a relaxation model by rewriting  all the couplings in terms of 3-form fields. The parameter $g$ of relaxation could be safely argued to remain naturally small even for a large field excursion of the axion.

\section{A minimal  relaxion monodromy model}
\label{sec:toy-model}

As mentioned in the introduction,  monodromic or multi-branched potentials have been suggested as a way out of certain puzzling features of the naive relaxion models. However, there is actually no explicit relaxion model with a built-in monodromy structure in the literature. In this section we describe  
the simplest relaxation model with an axion monodromy structure. We also revise the issue of the smallness of $g$ in the context of monodromy. Clearly, the construction admits many generalizations, and the model in this section is proposed as a simple illustration. In this sense, we emphasize that the analysis in forthcoming sections is actually meant for general constructions, rather than the precise model in this section.

\subsection{Seesaw-like scales and stability}
\label{sec:seesaw}

The scales in axion monodromy show an interesting see-saw structure which has not been pointed out in the literature.

As discussed in the previous section, requiring that an axion potential has a monodromic structure amounts to requiring that it is invariant under a discrete axion shift symmetry which acts non-trivially on the parameters of the potential. In the previous section we saw in \eqref{qmf} that 
\beq 
 g\ =\ k\frac {\Lambda_k^2}{2\pi f} \ .\label{seesaw}
\eeq
So we see that $g$ is quantized in units of $\Lambda_k^2/(2\pi f)$.  The structure in \eq{seesaw} is reminiscent of  the seesaw mechanism for neutrino masses (for a review, see e.g. \cite{King:2003jb}), where the very small neutrino mass arises as a quotient $m_W^2/M_\nu$ between two different mass scales; as long as $M_\nu\gg m_W$ (typically $M_\nu\sim 10^{10}-10^{16}$ GeV in the neutrino context), the resulting neutrino mass will be much lower than both $m_W$, $M_\nu$.

Similarly, (\ref{seesaw}) can explain one of the troubling aspects of the relaxation mechanism, namely the tiny value of $g$. This is typically required to be  as small as e.g. $10^{-34}$ GeV, so that, even if relaxation works, we seemingly must introduce a new energy scale far smaller than any other in an unmotivated way. However, the smallness of $g$ is nicely explained by the seesaw mechanism in \eq{seesaw}. For $f\simeq 10^{10}$ GeV one obtains for the 
flux scale $\Lambda_k$:
\beq
g=10^{-34} \ {\rm GeV} \longrightarrow \  \Lambda_k \ \simeq \ \sqrt{gf} \ \simeq \ 10^{-3}\ {\rm eV} \ .\label{relcop}
\eeq
So for typical relaxionic values the new fundamental scale is rather $\Lambda_k\simeq 10^{-3}$ eV, and  the smallness of $g$ appears as a derived very small quantity determined by the effective see-saw relation  $g\simeq \Lambda_k^2/f$. This beautifully follows from the monodromy version
of the axion symmetry, via  the discrete shift symmetry and the quantization of $f_0$. The new fundamental scale $\Lambda_k$ is much larger, 20 orders of magnitude larger than $g$.  

Before moving on we should notice an interesting numerical coincidence in this minimal example. 
The new scale $\Lambda_k\simeq 10^{-3}$ eV is in the order of magnitude to the observed cosmological dark energy scale  $\Lambda_{\rm dark}\simeq (10^{-3} {\rm eV})^4$. It is tantalizing to speculate that  both scales are physically related, perhaps through the intimate relation between (free) 4-forms and their contributions to the cosmological constant \cite{Brown:1988kg,Duncan:1989ug,Feng:2000if,Bousso:2000xa}. 

In any case, the smallness of $\Lambda_k$ is yet to be explained. A perhaps interesting observation\footnote{We thank the anonymous referee for pointing this out to us.} along this line is that, by combining \eq{axcop} and \eq{relcop}, we arrive at
\begin{align}\Lambda_k\sim \frac{\Lambda_v^2}{M},\label{seesaw2}\end{align}
This  resembles  a further seesaw between the nonperturbative scale $\Lambda_v$ and the SM cutoff scale $M$. However, in absence of any direct coupling between QCD and $F_4$, \eq{seesaw2} remains accidental. Indeed, \eq{axcop} means that the nonperturbative barriers are able to stop the relaxion, and thus constitute a purely phenomenological requirement for the relaxion picture to work.

This see-saw structure can provide an explanation for the originally tiny value of $g$. But one should also address the question of the stability of this parameter, both at the classical level (taking into account non-negligible higher dimensional operators due to the large field excursion of the relaxion) and at the quantum level due to loop corrections. The former makes reference to the infinite tower of non-renormalizable operators which a priori become relevant when the field takes values larger than the cut-off of the theory (as required in relaxation), while the latter refers to the quantum stability of the classical Lagrangian. The argument for which $g$ is technically natural since it is associated to a symmetry breaking is not valid in the context of monodromy, or at least, the underlying protection is more subtle, because indeed the discrete shift symmetry of the axion remains unbroken at the level of the action (so $g$ is not associated to the breaking of the discrete shift symmetry). However, monodromy provides a new mechanism to guarantee the stability of the effective potential. First, as explained in the previous section, the gauge invariance of the 3-form shields the potential against non-renormalizable higher dimensional operators, implying that those should come as powers of the potential itself. Therefore they will remain subleading (due to the original smallness of $g$) even if the field excursion of the relaxion is bigger than the cut-off scale. Let us remark that  the stability of the full scalar potential can only be guaranteed if the complete perturbative potential for the axion (including mass terms and interactions) arises from a coupling to a single 3-form field. Therefore, not only $V(g\phi)$ has to be rewritten in terms of a coupling to a 3-form field, but also the axion-Higgs coupling term. This will be the subject of section \ref{sec:coupling-multibranched}. 
In addition, the stability of the potential at quantum level appears as a natural consequence of the previous argument, because all the classical perturbative couplings involving the relaxion $\phi$ are then controlled by $g$. Therefore the quantum corrections will only give rise to a renormalization of the parameter proportional to itself, implying that $g$ is technically natural and will remain small if was originally small (e.g., due to the see-saw structure described above). 

\subsection{Coupling a multi-branched axion to the SM}
\label{sec:coupling-multibranched}

Let us start by recalling the  minimal relaxion model (\ref{relaxation}),
\beqa
V=V(g\phi)+(-M^2+g\phi)|h|^2+\Lambda^4\cos\left(\frac{\phi}{f}\right).
\eeqa

The simplest option for  a  monodromy  invariant version of this coupling one could think of is
\beq
V \ =\ V_{SM} \ +\ V_{KS} \ - \eta F_4|H|^2 \ +\ V_{cos}
\eeq
with $\eta$ some constant of order one and
\beq
V_{KS}= \frac12|F_4|^2-g\phi F_4
\eeq
One obtains, after eliminating $F_4$
\beqa
V\  &=&  \  -\mu^2|H|^2\ +\ \lambda|H|^4 \ +\ (f_0\ + \ g\phi \ +\ \eta |H|^2)^2 \  +\ V_{cos}=\\
&=& \ {\tilde \lambda }|H|^4 \ +  \ (f_0\ +\ g\phi)^2 
\ + \ 2\eta (-M^2 \ +\ g\phi)|H|^2 \ +\ V_{cos} ,
\eeqa
where we define
\beq
M^2\ =\ \frac{\mu^2}{2\eta} \ -\  f_0 \ .
\eeq
An important point to remark is that $\mu$ is of order the UV cut-off, includes all loops, and is not quantized in general.
Then $f_0$ is not required to be large, and no enormous  quanta are required.   It only has to shift appropriately 
and will generically be of order $f_0\simeq \Lambda_k^2$.  

The structure just obtained is similar to the relaxion model, with a certain KS-like potential for the axion.  Now, as the axion rolls down,
it meets first the point  at which $\phi=M^2/g$ as usual. But still goes down because the KS potential has not yet reached 
its minimum. But then eventually the cosine piece enters into the game and stops the relaxation, as usual.

This  minimal monodromy version of relaxation improves the original in several  respects. The gauge shift symmetry is preserved,
even though the axion has a non-trivial quadratic potential. This symmetry also protects the relaxion from Planck suppressed
corrections which appear in powers of $F_4^2/M^4$. Furthermore, the minute mass scale $g$ appears as derived
from a much larger fundamental scale $\Lambda_k \simeq 10^{-3}$ eV via the relationship  $g\simeq \Lambda_k^2/f$.

It is clear that many other models may be constructed by introducing different couplings of the 4-form strength to the Higgs field.
Also the values of the mass scales involved, the cut-off $M$ and the 4-form scale $\Lambda_k$ may be very different from the
ones in this simple example. As an example,  in section \ref{sec:string-relaxion} we show a model derived from a string setting in which the
coupling  of $F_4$ to the Higgs system is quite different.

We refrain from entering a detailed  model-building search,  and instead turn to the interesting question of whether the general idea of a monodromy relaxion is viable, by studying new model-independent constraints. In particular, the multi-branched structure of the axion potential implies the existence of tunnelling between vacua corresponding to different axion branches. We have to explore whether 
this tunnelling is sufficiently suppressed so that the relaxion indeed proceeds smoothly through slow roll to reach 
the point in which it induces a massless Higgs. We study these issues in the next two sections, in particular exploiting the implications of the Weak Gravity Conjecture.

\section{Membrane nucleation and the Weak Gravity Conjecture}\label{WGCmon}

Axion monodromy models, in particular those involving a large number of axion windings along a branch, must address the question of possible tunneling transitions between branches (mediated by membrane nucleation) which can reduce the effective field range of the axion (see e.g. \cite{Kaloper:2008fb,Kaloper:2014zba,Marchesano:2014mla,Franco:2014hsa} for discussion in the axion inflation setup). However a quantitative estimate of the tunneling probability requires information from the UV completion, which determines the tension of the corresponding membrane. As we argue in Appendix \ref{WGC3}, in theories containing quantum gravity, a version of the Weak Gravity Conjecture (WGC) in \cite{ArkaniHamed:2006dz} can provide useful information on the parametric dependence of the membrane tensions, in a fairly model-independent way, and therefore can yield generic constraints on relaxion models. In this section we study the membrane nucleation process 
and explore possible constraints from the WGC.

\subsection{Membranes and monodromy}\label{sec:membranes}

In a generic relaxion model, an immediate question is how large a hierarchy we can obtain between $\Delta\phi$, the field range traversed by the relaxion during inflation, and $f$, the relaxion decay constant. Can one really restrict to a single branch and go up the potential to make the field range  parametrically large? The answer is no, in general. As is well-known in the monodromy literature \cite{Marchesano:2014mla}, in the presence of membranes there are dynamical processes which make the field jump from one branch to a lower one, spoiling the slow rolling
\footnote{There are other processes which may spoil too large windings of the axion, see e.g. \cite{Kaloper:2014zba}. In this paper we restrict to membrane nucleation, which as we show is enough to stress the relaxion a little bit.}. However, as for any non-perturbative tunneling process, the probability for this to happen is in principle exponentially suppressed (see \cite{Kaloper:2011jz, Franco:2014hsa} for some discussion in the context of axion monodromy inflation). 

Let us describe this process in more detail. During inflation, and the rolling of the relaxion, a bubble may nucleate, bounded by a membrane. The value $f_0$ of the $4$-form will jump by $\Lambda_k^2$ upon crossing the membrane.  Since the vacuum energy is lower within the bubble than outside, the bubble will expand indefinitely, provided that it is initially large enough so that the pressure associated to the difference in vacuum energies beats the surface tension. The smallest bubble for which this happens is the so-called critical bubble.

Bubbles smaller than the critical radius cost energy to produce, since the surface tension overcomes volume. As a result, they cannot be produced in the vacuum. The critical radius bubble costs no energy, and hence it can be produced by an instanton effect. One can estimate the transition rate for this process in the thin wall approximation. There is a well known expression for this
\beq
P\sim \exp\left( -B\right),\quad  B=\frac{27\pi^2 T^4}{2(\Delta V)^3} \label{decayrate0}
\eeq
where $T$ is the tension of the membrane which induces a shift on $f_0$ and $\Delta V$ is the variation in the potential energy, in our case 
\beq
\Delta V=V_i-V_f\sim \Lambda_k^2g\phi.
\eeq
This formula however neglects gravitational effects. As discussed in \cite{Coleman:1980aw}, it is only valid when the gravitational backreaction of the energy density in the bubble can be ignored. This will be the case whenever the bubble radius\footnote{Notice that $r\sim T/(\Delta V)$ is the radius that the bubble would have in flat space. When gravitational effects are important, the expression for the bubble radius is modified, as explained in Appendix \ref{sec:bubbles}.} $r\sim T/(\Delta V)$ is smaller than the de Sitter radius $H^{-1}$ associated to the energy density of the bubble. In other words, gravitational effects are negligible if
\begin{align}q\equiv \frac{1}{rH}\sim \frac{\Delta V}{TH}>1\label{q}\end{align}
where the variable $q$ parametrizes the importance of the gravitational effects and will be useful later.
We will see in section \ref{sec:WGCconstraints} that  relaxionic models correspond to precisely the opposite regime.  For typical parameters, $q=(rH)^{-1} \leq1$ and gravitational effects are significant. 

We therefore need to use the more general expression for vacuum decay which can be found in  \cite{Coleman:1980aw}, see Appendix \ref{sec:bubbles} for details. It turns out that the approximate expression valid in the relaxation regime is 
\begin{align}B\approx w(q)\frac{2 \pi ^2 T}{H^3}.\label{decayrate}\end{align}
Here, $w$ is a certain function of $q$ defined in Appendix \ref{sec:bubbles} and ranging from one to $\approx 0.1$ in the relaxation regime.
Notice that unlike \eqref{decayrate0}, \eqref{decayrate} depends strongly on the Hubble constant $H$, signalling the importance of gravitational effects. 

Formulae like \eqref{decayrate0} or \eqref{decayrate} are useless without additional information, because they give the tunneling probability in terms of the membrane tension $T$, which cannot be constrained from just effective field theory. It is here where the WGC  can be helpful, allowing us to constrain the model even if we do not know the exact UV completion. If the WGC predicts the existence of a membrane with very small $T$, we have $B\ll1$. Even though we would be outside of the semiclassical approximation inherent to any instanton computation, a small value of $B$ would generically mean that membranes  are produced copiously, since their nucleation is not suppressed by exponential effects. Since inflation lasts for so long in relaxation models, we would produce enough membranes so as to completely spoil the slow-roll of the relaxion. Generically, if a strong enough form of the WGC holds, we can use it as  a tool to discern which effective field theories might be a priori UV completed when including gravity and which ones would be in the Swampland instead.

\subsection{WGC and membranes}\label{WGCmembranes}
We now briefly discuss what the Weak Gravity Conjecture is and apply it to 3-form fields. This conjecture has proved useful in constraining inflationary models based on natural inflation with one or multiple axions \cite{delaFuente:2014aca,Rudelius:2014wla,Rudelius:2015xta,Montero:2015ofa,Brown:2015iha,Bachlechner:2015qja,Hebecker:2015rya,Brown:2015lia,Junghans:2015hba,Heidenreich:2015wga,Palti:2015xra,Heidenreich:2015nta,Kooner:2015rza,Kappl:2015esy,Choi:2015aem}. Our analysis is the first dedicated study of WGC constraints to axion monodromy models as well\footnote{One could also try to apply this conjecture to axion monodromy models in inflation (see \cite{Brown:2015iha} for partial attempts). However, the constrains derived in that case are too weak and do not pose a serious problem for inflation. It is the particular interplay between the scales in relaxation what makes these constraints relevant.}.

The original statement of the WGC \cite{ArkaniHamed:2006dz} is  that, in theories of quantum gravity, for an abelian $p$-form gauge field with coupling $g_p$, there must exist a charged object, of charge $Q$ and $p$-dimensional worldvolume, which is superextremal, i.e. its tension $T$ satisfies
\begin{align}T\lesssim\frac{g_pQ}{\sqrt{G_N}}\label{WGCmild}\end{align}
where $G_N$ is Newton's constant. 

Such state is required to allow the decay of subextremal charged objects which can be constructed in the effective field theory of the $p$-form field coupled to gravity. For instance, for $p=1$ in $d=4$, this effective theory contains subextremal Reissner-Nordstrom black holes. These black holes may lose charge and mass via Hawking radiation of charged particles, and become extremal. However, extremal black holes risk becoming superextremal when radiating charged particles. Superextremal black holes are problematic, since they have naked singularities which generically drive the theory out of control. So, either they are stable, or there is some decay channel which preserves (sub)extremality. The former possibility is not feasible, at least in theories (like string theory) in which the coupling $g_p$ is tunable: by taking $g_p$ very small we could get a large number of stable, almost massless and degenerate, extremal black holes,  resulting in the usual trouble with remnants \cite{Susskind:1995da}, as explained in \cite{Banks:2006mm}. 

We are thus let to the latter possibility. The requirement for an extremal black hole with  charge $Q$ and mass $M=Q\,g_1M_p$ to be able to decay while remaining (sub)extremal is that there is a particle with mass $m$ and charge $Q_m$ such that
\begin{align}
\frac{(Q-Q_m)\,g_1M_P}{M-m}\leqq1\quad \longrightarrow \quad Q_m\,\frac{g_1M_P}{m}\geq1
\end{align}
where we have assumed  a black hole arbitrarily close to extremality, $M\sim Q g_1 M_p$. This is precisely \eq{WGCmild} for $p=1$; a similar argument works for (almost) any other $p$.

In this paper we are concerned with applying the WGC to $3$-forms in four dimensions. This is a subtle issue, which we discuss in apprendix \ref{WGC3}, but our conclusion is that the WGC is plausible (up to order 1 factors) in this case as well. Furthermore, as discussed in the appendix, we will assume a \emph{strong} form of the WGC, which basically says that one may take $Q=1$ in \eq{WGCmild} if desired. This is a particular instance of the ``Lattice Weak Gravity Conjecture'' of \cite{Heidenreich:2015nta}, which is the only strong form of the conjecture known so far that is consistent with dimensional reduction. Although the black hole arguments above only motivate the mild form of the conjecture, no counterexamples to the strong form are known in string theory. In fact, the membranes predicted by the WGC are present in any stringy model we can think.

In terms of 3-form data, the gauge coupling $g_3$ is none other than $\Lambda_k^2$, the 4-form flux quantum, which as argued in section \ref{sec:axion-monodromy} equals $2\pi fg$ in monodromic relaxion models.  The strong form of the WGC applied to 3-form fields in four dimensions implies that the membrane which shifts the flux $f_0$ by one unit must have a tension lower than
\beq
T\sim 2\pi fg M_P
\eeq
where we have restored the Planck mass,  $f$ is the decay constant of the axion, and there might be additional $\mathcal{O}(1)$ factors. The tunneling probability associated to membrane nucleation is given by eq. \eqref{decayrate} 
\beq
B\approx 4\pi^3w(q)\frac{fg M_P}{H^3}. 
\label{prob}
\eeq

One can now take a particular effective relaxation model  and compute the tunneling probability between different branches.  If the tunneling is not suppressed, many such transitions will take place over the exceedingly large number of e-folds that inflation lasts in the original relaxion proposal, spoiling the simple slow roll picture of the relaxion dynamics.

This mechanism will not stop with the appearance of the QCD barriers, since periodic potentials coming from nonperturbative effects have (approximately) the same value at $\phi$ and $\phi+2\pi f$; only the monodromic part of the energy changes. As a result, $\Delta V$ remains the same and so does the transition rate. 

It is important to emphasize the WGC membrane is a kind of domain wall different from the field-theoretic ``bounces'' which arise once the QCD barriers switch on. These are fully accounted for by low-energy effective field theory, have a tension $T\sim f\Lambda_v^2$ (where $\Lambda_v$ is of order of the strong coupling scale of the gauge group coupled to the relaxion) in relaxionic models, and are  already discussed in \cite{Graham:2015cka}. Their decay rate \cite{Coleman:1977py,Coleman:1980aw} can be safely computed within effective field theory and is not an issue for the relaxion proposal. On the other hand, the effect of the membranes we consider cannot be computed within effective field theory, and we need additional input like the explicit value of $T$, or the WGC bounds on it.

\section{Constraints on the relaxion}\label{sec:killrel}

When studying the viability of relaxation in a UV completion using monodromy there are two issues to address. First, the stability of the full scalar potential can only be guaranteed if the complete potential for the relaxion (including also the axion-Higgs coupling term) is rewritten in terms of a coupling to a 3-form field. This has been the subject of section \ref{sec:toy-model} from a pure effective field theory point of view. First attempts to find such a structure in string theory will be given in section \ref{sec:string-relaxion}.

Secondly, if one indeed succeeds in constructing such a model, one should study the tunneling probability by membrane nucleation before making any claim about the effective field range available to relaxation.

A priori it makes no sense to address the second issue without having a complete realisation of monodromic relaxation, since one needs the tension of the membrane (and more generally, the action of the instanton) to be able to make any concrete claim. However, it turns out that the typical scales involved in the theory to address the EW hierarchy problem are in general so extreme that we can already draw some conclusions just focusing on the pure relaxion potential $V(g\phi)$. Taking $g$ to be the 3-form coupling, one can apply the formulae in the previous section to  constrain specific relaxion models.

\subsection{Constraints on the relaxion parameter space}
The original  relaxion model \cite{Graham:2015cka} has a parameter space specified by the cutoff $M$, the axion coupling $g$,  the axion decay constant $f$,  the Hubble scale $H$ during inflation, and the energy scale $\Lambda$ of the gauge group providing the nonperturbative effects which stop the rolling of the potential. If we want the relaxion to also solve the strong CP problem, then $f$ and $\Lambda$ are constrained to have their QCD values. This parameter space is however very constrained. We now review these constraints as established  in \cite{Graham:2015cka}.

In order for the relaxation mechanism to provide a dynamical solution to the EW hierarchy problem for generic initial conditions of the relaxion $\phi$, inflation must last long enough for $\phi$ to scan the entire range of the Higgs mass. This implies that 
\beq
\Delta \phi\gtrsim M^2/g
\label{dphi}
\eeq
leading to a lower bound in the number of efolds 
\begin{align}N\gtrsim\frac{H^2}{g^2}\end{align}
about $N_e> 10^{37}-10^{67}$, depending on the specific details of the model and further constrains on inflation. The relaxion will stop rolling when the barrier of the non-perturbative potential becomes comparable to the slope of the perturbative potential, ie.
\beq
gM^ 2\sim \frac{\Lambda^4_v}{f}
\label{g}
\eeq
where $\Lambda^4_v\equiv\Lambda^4(h=v)=cv^2$ being $c=f_{\pi}^2y_u^2$ if $\phi$ is the QCD axion.  If we require $M\gg m_W$ (to have a EW hierarchy problem to solve), the relation \eqref{g} implies an upper bound
\beq
g\ll \frac{\Lambda_v^4}{fm_W^2}
\label{gll}
\eeq
For the minimal field content in which the relaxion is the QCD axion, the above bound reads $g\ll 10^{-16}$ GeV, where we have used $\Lambda_v\sim \Lambda_{QCD}\sim 200$ MeV and $f\sim 10^9$ GeV. For instance, a cutoff $M\sim 10^7$ GeV implies $g\sim 10^{-26}$ GeV. Combining eqs.\eqref{dphi} and \eqref{g} we get
\beq
\frac{\Delta\phi}{f}\gg \frac{M^4}{\Lambda_v^ 4}\sim \frac{M^4}{cv^2}
\eeq
Therefore a big hierarchy between the EW scale and the cutoff of the theory implies in turn an even bigger hierarchy between the axionic field range and the fundamental periodicity $f$. Notice that this is independent of whether the relaxion is the QCD axion. In terms of the monodromic model, this implies that the axion must travel its fundamental domain an extremely huge number of times, between at least $10^4-10^{40}$ times for a cutoff $M\sim 10^4-10^{13}$ GeV. It is then reasonable to expect a non-negligible tunneling probability  so that it is more efficient to decrease the energy by jumping from one branch to another than by slowly rolling down the potential. In the next section, we explicitly compute this probability by plugging the above constraints in the formulae derived for the transition rate in the previous section. But before that, let us recall that the Hubble constant is also highly constrained in the relaxion models.
For the QCD-like barriers to form, we must have
\begin{align}H<\Lambda_v.\label{barriers-form}\end{align}
Also, energy density during inflation must be dominated by the inflaton, rather than the relaxion. As a result
\begin{align}H>\frac{M^2}{M_P}\label{energydens}.\end{align}
This already imposes an upper bound on the cutoff given by $M\lesssim(\Lambda_v M_p)^{1/2} \sim 10^{9}$ GeV for $\Lambda_v$ taking the QCD value. This constraint can even be a bit stronger if we also impose that quantum fluctuactions are subleading with respect to the classical rolling, leading to $M\lesssim 10^{7}$ GeV .

\subsection{WGC constraints\label{sec:WGCconstraints}}
The above constraints were already discussed in the original relaxion paper.  We want to study whether 
membrane nucleation provides  extra constraints  on the relaxion parameter space.

If the transition rate is not exponentially suppressed, we can expect a large number of membranes being produced during inflation. If this number is larger than the number of times the relaxion winds its fundamental domain, $\approx 10^4-10^{40}$, membranes will efficiently take $\phi$ to 0, thus spoiling the solution to the hierarchy problem. Therefore, in order to have successful relaxation we must have $P\sim \exp(-B)\ll 1 $.

Inserting the relations between $f,g,H$ and $M$ found in the previous section (given by eq.\eqref{g}) into \eqref{prob} and using $\phi\sim M^2/g$, we get
\beq
B\approx 4\pi^3 w(q)q^3 \frac{\Lambda_v^4 M_P}{M^2}\label{prob1-2}
\eeq
The variable $q$ was defined in \eqref{q} and parametrizes the importance of the gravitational effects. If $q=1$ we recover the flat-space formula \eqref{decayrate0}. In particular, we have
\beq
q=\frac{\Delta V}{HT} =\frac{g\phi}{HM_p}\sim \left(\frac{\phi}{M^2/g}\right) \frac{M^2/M_p}{H}<1\label{q22}
\eeq
which is always smaller than one at the end of relaxation, since for relaxion models satisfying $\phi\sim M^2/g$, the value of $q$ is controlled by the ratio between the density energy for relaxion and inflation. Therefore gravitational effects may play a role suppressing the value of $B$ in \eqref{prob2} and increasing the tunneling probability. Since $H<\Lambda_{v}$ (see \eqref{barriers-form}) we also have a lower bound for $q$ given by 
\beq
\frac{M^2}{M_p\Lambda_v}<q<1.
\label{qbound}
\eeq
Using \eq{q22}, and taking into account $\phi\sim M^2/g$ during relaxation, we may  rewrite \eq{prob1-2} as
\beq
B\approx 4\pi^3 w(q)q^3 \left(\frac{\Lambda_v M_P}{M^2}\right)^4\label{prob2}.
\eeq

With this expression, we can compute the total number of bubbles nucleated in a region of spacetime volume $V$ simply as
\begin{align}N_b\sim V Pe^{-B},\end{align}
where $P$ is a model-dependent instanton prefactor which typically is not exponentially suppressed. As derived in Appendix \ref{sec:bubbles}, once a bubble starts expanding it reaches the de Sitter radius in a time of order one Hubble scale. Since in relaxation inflation lasts for so many e-folds, this means that a given observer will be within virtually any bubble nucleated in his/her past light cone. For each bubble we are in at the end of relaxation, the relaxion jumps by $2\pi f$ towards the minimum of the potential. As discussed above, too many of these jumps have the potential to drive the relaxion all the way to its minimum, thereby spoiling the relaxation mechanism. Hence we should compute $N_b$ taking $V$ as the volume of the past light cone of a point at the end of relaxation. Nevertheless, we cannot give a precise value for $N_b$, since the instanton prefactor $P$ is model-dependent and generically difficult to compute. We will instead obtain constraints by demanding that the exponential suppression $e^{-B}\ll1$. This is a sufficient condition to guarantee that bubble nucleation does not spoil relaxation.

Requiring a suppressed tunneling probability with $P\sim \exp(-B)\ll 1 $, yields  a constraint
 \begin{align}M\leq (4\pi^2w(q)q^3)^{1/8}\sqrt{\Lambda_v M_P}\sim (w(q)q^3)^{1/8}\ 2.5\cdot 10^9\ GeV\label{upbound}\end{align}
 where we have used $\Lambda_v\approx \Lambda_{QCD}$.
 This constraint for $M$, based only on the tunnelling rate,  is slightly weaker than
those already discussed in \cite{Graham:2015cka}, unless $q\ll1$. However, it can be checked that due to the lower bound for $q$ in \eqref{qbound} it is not possible to get a stronger constraint for the cutoff than $M<10^9$ GeV.

 Let us remark that the only constraints coming from the relaxion proposal that we have used are: the fact that $\phi \sim M^2/g$ (necessary to get a light effective Higgs mass), the relation \eqref{g} derived from imposing that the non-perturbative barrier eventually stops the running of $\phi$, and the requirement that the energy density is dominated by inflation instead of relaxion, leading to \eqref{energydens}.  Hence the result is quite independent of the specific relaxion proposal, as long as it satisfies the three previous conditions and has a built-in monodromy structure.

For instance, the minimal relaxion model described in section \ref{sec:toy-model} includes a new parameter $\eta$ which measures the coupling of the Higgs field to the Minkowski 4-form. The constraints for relaxation are modified so that the bound in the cut-off becomes
\begin{align} \mu\lesssim (4\pi^2w(q)q^3)^{1/8}\sqrt{\Lambda_v M_P\eta}.\label{crexeta}\end{align}
One could a priori think of relaxing the constraint on the cut-off by having an extremely large $\eta$.  However this leads the theory out of perturbative control, since $\eta^2$ is a coefficient for the Higgs quartic coupling.

\begin{figure}[t]
\begin{center}
\includegraphics[width=6.2cm]{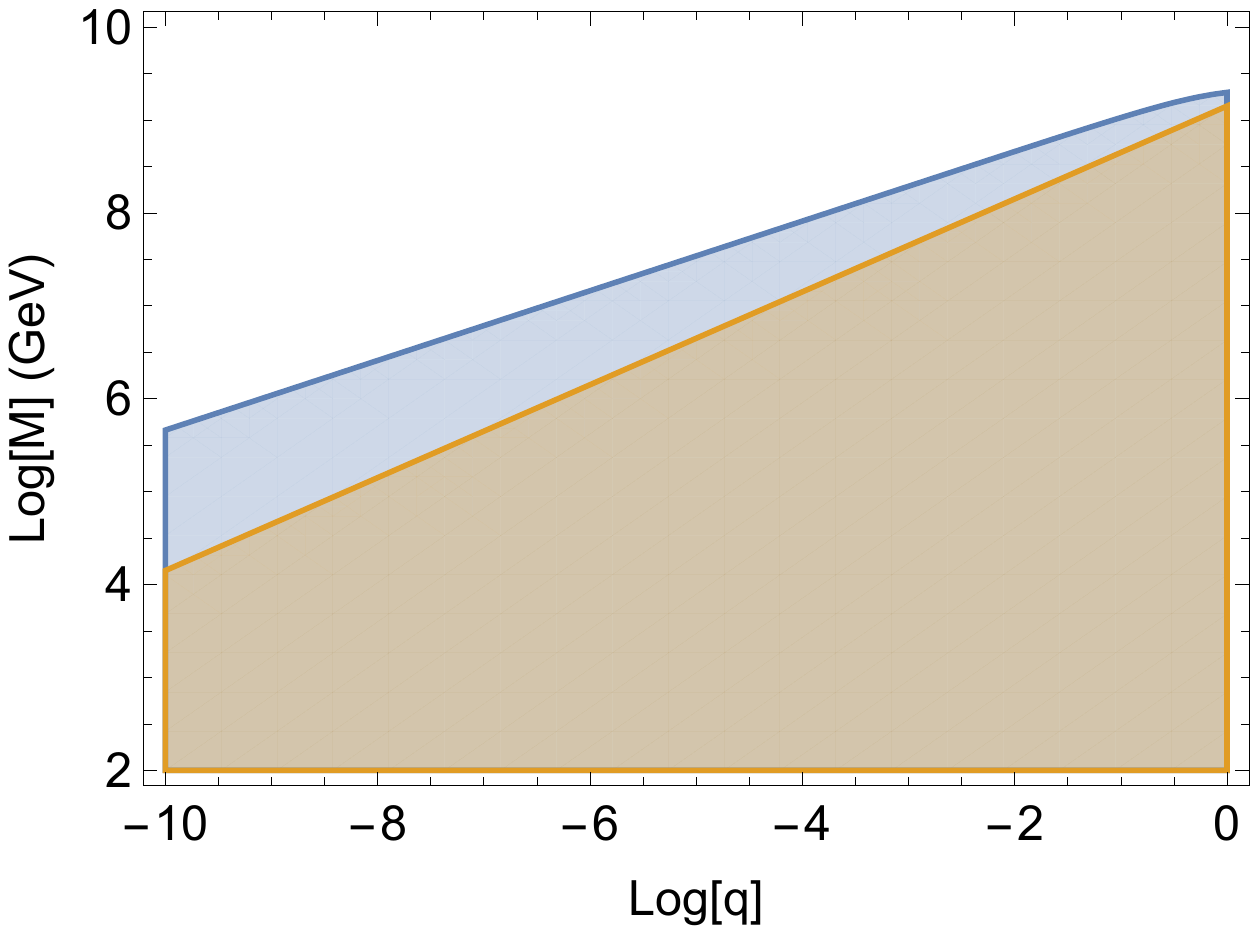}
\includegraphics[width=8.4cm]{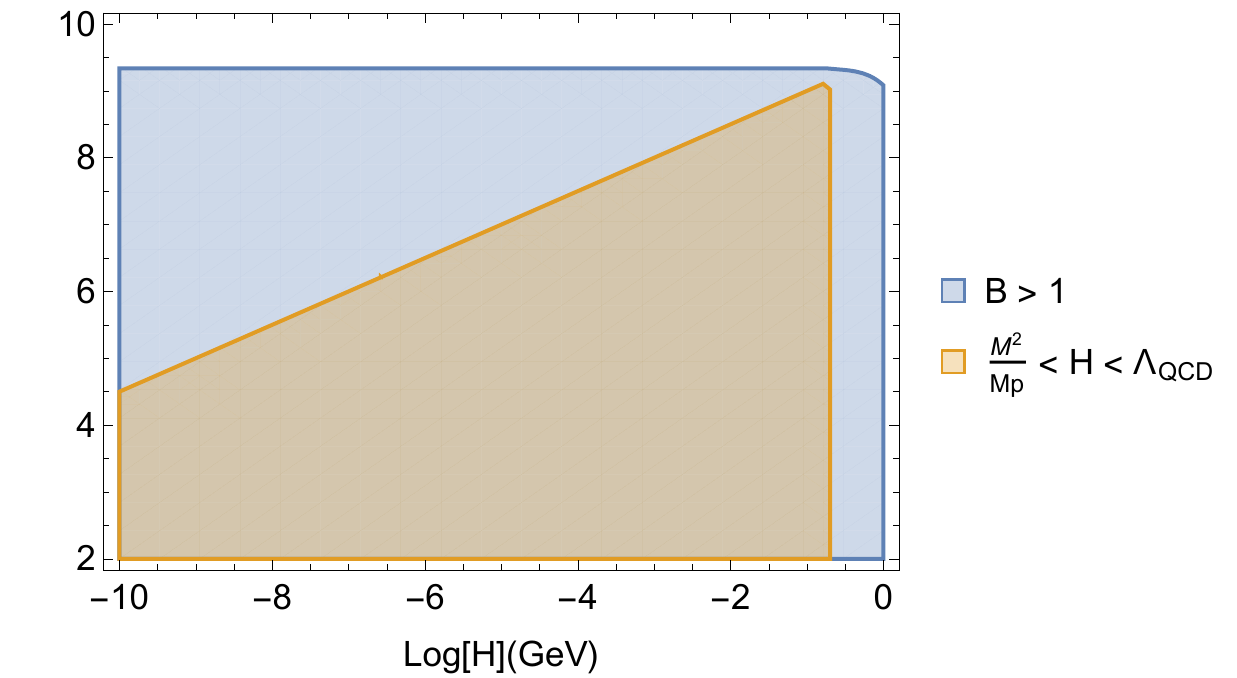}
\end{center}
\caption{Plot of constraint $B>1$ including the $w(q)$ dependence, in blue, and of the constraints coming from coming from eqs.\eqref{barriers-form} and \eqref{energydens}, plotted in orange. The blue area corresponds to the region surviving to the tunneling constraints in the parameter space spanned by the cut-off scale $M$ and the parameter $q\simeq M^2/(HM_p)$ (left) or the Hubble scale (right).}
\label{constris}
\end{figure}

 In Figure \ref{constris} (right) we plot the dependence of the transition rate on the cut-off scale $M$ and the Hubble scale $H$ (which enters in the formulae through the dependence on $q$). In the left Figure we plot the constraints directly in terms of $q$, to show explicetely the relevance of the gravitational effects. The blue region corresponds to the allowed regime with $B>1$. The constraints on the Hubble scale (yielding \eqref{qbound}) put an additional constraint on the cut-off,  represented by the orange region. This latter constraint is the one already appearing in \cite{Graham:2015cka}. As can be seen in the picture, the $B>1$ and these constraints are similar along the relaxion parameter space. However, their physical origin is completely different. The $B>1$ constraints constitute yet another example of the use of the Weak Gravity Conjecture to constrain interesting effective field theories.
  
 Let us also finally comment that the minimal version of cosmological relaxation is already ruled out by the strong CP problem. The minimization of the axionic potential implies
\beq
\theta_{QCD}\sim \langle \phi\rangle \sim gM^ 2f^2/\Lambda_v^4 \sim \mathcal{O}(f)
\eeq
while the experimental constraint imposes $\theta_{QCD}< 10^{-10}$. There are mainly two proposals to solve this problem still keeping the relaxion to be the QCD axion. Either the slope of the perturbative axionic potential decreases dynamically after inflation \cite{Graham:2015cka}, or the non-perturbative potential induced by QCD instantons is suppressed during relaxation and grow afterwards to force  $\theta_{QCD}$ to small values \cite{Batell:2015fma}. In the former case, the constraint \eq{upbound} is modified to
 \begin{align} M\lesssim (4\pi^2w(q)q^3)^{1/8}\sqrt{\Lambda_v M_P}\theta^{\frac{5}{16}}\approx1870\ TeV.\label{crex2}\end{align}
 Again, the order of magnitude of these bounds is higher than the ones obtained from entirely different arguments in \cite{Graham:2015cka}. Nevertheless, the tunneling constraints are completely independent and provide an example of somewhat stringent quantum gravity constraints on effective field theories. We stress however that this bounds only apply to the particular solution of the strong CP problem in \cite{Graham:2015cka}; other solutions, such as the one in \cite{Batell:2015fma}, might still be viable.

The third option is to consider an additional new strong group coupled to the Higgs which stops $\phi$ from rolling. The scale $\Lambda_v$ can be then increased until a few hundred GeV \cite{Graham:2015cka}, so the tunneling probability is further suppressed and the constraint for the cut-off is relaxed to  $M\lesssim 10^8$ GeV, again times the cosmological suppression factor $(w(q) q^5)^{1/16}$. These latter constraints are comparable (perhaps slightly stronger) to the ones obtained in \cite{Graham:2015cka}.

\section{Monodromy relaxions and string theory}\label{sec:string-relaxion}

Many of the ingredients of the relaxion models discussed in previous sections are present in string theory. 
Periodic axion-like fields appear in all string compactifications, and their shift symmetries are  typically remnants 
of gauge invariance of higher dimensional antisymmetric fields (see \cite{Ibanez:2012zz} for review). In particular in Type II orientifolds axions appear from
expanding  D=10 Ramond-Ramond (RR) antisymmetric fields $C_{MN..Q}$ on harmonics over the 6D compact space, see e.g.\cite{Grimm:2007hs}.
They also appear upon dimensional reduction of Neveu-Schwarz(NS) antisymmetric 2-forms $B_{MN}$.  
The field strengths of these antisymmetric fields as well as the magnetic fluxes of $F_{MN}$ regular gauge fields are
quantized when integrated over the compact space. All these integer quanta are inherent discrete degrees of freedom
for any string compactification.  This leads to a {\it landscape} of string vacua for any given particular compact space, which has
been argued to  be at the root of the understanding of the smallness of the value of the cosmological constant \cite{Bousso:2000xa}.

As we mentioned above, it has been realised that in large classes of Type IIA/IIB string compactifications the axions may have non-trivial
perturbative scalar potentials without spoiling the gauge discrete shift symmetries. This is true as long as not only the axion gets shifted, but
also the (quantized flux) parameters  of the potential transforms appropriately. The structure is similar to the Kaloper-Sorbo type of
potential discussed above generalised to include multiple axions and higher  order polynomial interactions.  Thus the structure is again that of monodromic axions,
whose symmetries are again better described in terms of 4d 3-forms, as in the simple examples discussed in this paper,
see  \cite{Bielleman:2015ina} for a more general discussion.

So string theory contains two of the required ingredients to construct relaxion models: 1) there are axions  and 2) they have 
the required multi-branched structure so that the axion potential does not spoil the shift symmetry.
In this section we try to take a further step and attempt to build a string construction with the  required
axion-Higgs couplings appearing in a relaxion model.  We use Type IIB orientifolds, which have a rich  structure of axions, and exploit D-brane physics to  engineer the monodromy in an eventually  semirealistic construction. We will just provide an  explicit example of the required structure within a toy model, and will not pursue the building a complete model.  We will also see the limitations of the approach and the prospects for a more realistic structure. The reader not familiar with string technicalities may jump safely to eq.(\ref{stringrelaxion}) where the main example of this section is provided.

The setup we consider is a stack of 3 parallel D5-branes wrapping a 2-cycle in a Type IIB orientifold compactification\footnote{See chapters  11 and 12  in \cite{Ibanez:2012zz} for a detailed discussion of this class of string vacua.} with an orientifold with  O3/O7-planes.  In this type of compactifications there are axion fields $b_a$ arising from the expansion of the NSNS field $B_{MN}$ on harmonic 2-forms $\omega_a$ which are odd under the orientifold reflection. We choose our axion $\phi$ to be one of this kind, $B_2=\phi \,\omega_2$. This structure is similar to the one appearing in the original axion monodromy models of string theory in  \cite{Silverstein:2008sg,McAllister:2008hb}.

Let us now describe the realization of the SM gauge and Higgs fields. The initial gauge group associated to the three  D5-branes is $U(3)$, and in general there are adjoint scalar fields $\Phi_i$ which parametrize the motion of the D5-branes in the  four compact dimensions transverse to the branes. The dynamics of these scalars is described by the Dirac-Born-Infeld (DBI)  plus Chern-Simons action, as we discuss in appendix \ref{sec:DBI}. After some simplifications the DBI action applied to the mentioned system has the structure
\beq
S_{DBI}=-\mu_5 g_s^{-1}{\rm Tr}\int d^6\xi \sqrt{ \left(1+2\sigma^2\partial_\mu\Phi^i\partial^\mu\bar\Phi^i\right)\left(1+\frac12g_s^{-1}\CF_{ab}\CF^{ab} \right)\left( 1-4\sigma^2|[\Phi^1,\Phi^2]|^2\right)^2} \ ,
\eeq
where $\mu_5= (\alpha')^{3}/(2\pi)^5$ is the D5-brane tension, $g_s$ is the string coupling (dilaton) and
\beq
\CF_{ab}=\sigma F_{ab}-B_{ab} \ , 
\eeq
where $\sigma = 2\pi \alpha'$ and $(\alpha')^{-1}$ is the string tension.
We allow for magnetic flux quanta along the $U(1)$ of the $U(3)$,  $F_2=q\,\omega_2$,  and also assume there is an axion zero mode $B_2=\phi \,\omega_2$.
Here $\omega_2$ is an odd 2-form Poincare dual to the 2-cycle $\Sigma_2$ wrapped by the $D5$'s. 
Expanding this expression to second order in 4d derivatives we obtain
\beq
S_{DBI}=-\mu_5 g_s^{-1}\text{STr}\int d^6\xi \left( 1-4\sigma^2|[\Phi^1,\Phi^2]|^2\right)\left(1+\sigma^2 \partial_\mu\Phi^i\partial^\mu\bar\Phi^i+\frac14 g_s^{-1}\CF_{ab}\CF^{ab}+\dots  \right)
\eeq
where we have neglected higher order terms in $\CF$. Now take  the adjoints $\Phi_1$, $\Phi_2$ as
\begin{align}\Phi^1&=\left(\begin{array}{ccc}0&0&h^1\\0&0&h^2\\(h^1)^*&(h^2)^*&0\end{array}\right),\quad \Phi^2=\left(\begin{array}{ccc}0&0&0\\0&0&0\\0&0&m\end{array}\right),\nonumber\\ [\Phi^1,\Phi^2]&=\left(\begin{array}{ccc}0&0&mh^1\\0&0&mh^2\\-m(h^1)^*&-m(h^2)^*&0\end{array}\right).\end{align}
This correspond to one of the 3 D5-branes getting slightly displaced from the rest,  giving rise to a configuration with a gauge group $SU(2)\times U(1)$. Here $m$ is generically of order
the string scale, i.e. $m^{-2}\simeq \alpha'$.
Now, the potential for the axion arises from 
\beq
\frac14 g_s^{-1}\CF_{ab}\CF^{ab}\propto   (\sigma F_2\ -\ B_2)^2 \propto  (q-{\tilde g}\phi)^2 \ .
\eeq
This potential may be understood as arising from a Kaloper-Sorbo structure \cite{Retolaza:2015sta}. The relevant 4-form $F_4$ is the dual of $F_2$ in the D5-brane worldvolume, and has a cross  coupling $F_4\wedge B_2$, which upon using the equations of motion for $F_4$ and $B_2=\tilde{g}\phi\omega$ produces the above expression.
On the other hand using the matrices for $\Phi_1,\Phi_2$ above  and tracing over gauge indices, we get
\beq
\Tr|[\Phi^1,\Phi^2]|^2=-m^2|h|^2
\eeq
leading to a scalar potential
\beq
V_{DBI}\ = \ \   \mu_5V_{\Sigma_2}g_s^{-1}\left( 1+4\sigma^2m^2|h|^2\right)\left(1+(q-{\tilde g}\phi)^2 \right) \ .
\eeq
There is a built-in KS symmetry which enforces  the dependence on the axion to necessarily appear in powers of $F_4=(q-{\tilde g}\phi)$. 
However, that is not the case for the Higgs field, so that a large mass term of order the cut-off  is expected to appear at the quantum level, beyond the 
original DBI contribution to its mass. Including this one  obtains 
\beq
V \ =\ -\mu^2|h|^2 \ +\ V_{DBI} \ =\   M^4(q\ -\ {\tilde g}\phi)^2 \ +\ (-\mu^2  \ +\ 4M^4\sigma^2 m^2(1+(q-{\tilde g}\phi)^2) \ |h|^2 \label{hmass}
\eeq
where we have assumed that the quantum correction of  the Higgs mass is negative and $M^4=\mu_5V_{\Sigma_2}g_s^{-1}$, and have ignored a constant term which is not relevant for the discussion.
The Higgs kinetic term  is not canonical, and after redefining to the canonically normalised Higgs ${\tilde h}$ one finally gets a potential
\beq
V \ = \  \left( - \frac {\mu^2}{1+(q-{\tilde g}\phi)^2} \ +\ 4M^4\sigma^2m^2\right) |{\tilde h}|^2 \ +\  M^4(q\ -\ {\tilde g}\phi)^2 \ .
\label{stringrelaxion}
\eeq
The structure of this potential is of relaxion type, although not of the minimal class we discuss in the text.  At large $\phi$ the Higgs mass-squared  is 
positive, the axion then starts decreasing and at a certain point the Higgs mass vanish, a vev develops and then the non-perturbative QCD 
(or other) condensate\footnote{This can arise from non-perturbative gauge dynamics on D-branes to which the axion couples, or from euclidean D-brane instanton effects, see \cite{Ibanez:2012zz} for background.} stops the vev.  Note that one can obtain the above potential if a 4-form $F_4$ in a KS term couples also 
to the Higgs field through a term in the action of the form 
$F_4^2/(1+4\sigma^2m^2|h|^2)$.

This is an interesting toy model with some similarities with the simplest relaxion model, but still far from a fully realistic realization. 
The most obvious difficulty in a string embedding of the original relaxion dynamics  comes from the required mass scales. The natural scale in the above potential is the string scale, whereas in  a relaxion model  like that in section 3 the 4-form scales are of order $10^{-3}$ eV. 
One could perhaps consider models with low string scale, but that would also lower the cut-off scale $M$. 

To see this in more detail, let us compute the relaxion couplings $M$, $g$ and $f$ which we used in previous sections in terms of stringy parameters. First, let us fix the axion decay constant. The IIB supergravity part of the action relevant here is  \cite{polchinski1998string,Ibanez:2012zz} 
\begin{align}-\frac{1}{4\kappa_{10}^2}\int d^{10}x\vert H_3\vert^2,\label{bfield10d}\end{align}
where $\kappa_{10}^2=\frac12(2\pi)^7\alpha'^4$. The $B$ field couples to worldsheet instantons wrapping $\Sigma_2$ \cite{polchinski1998string,Ibanez:2012zz} as
\begin{align}\frac{1}{2\pi\alpha'}\int_{\Sigma_2} B_2\label{axdecons}\end{align}
We now expand $B_2=\tilde{g}\phi \omega_2$, where we remind the reader that $\omega_2$ is the volume form of $\Sigma_2$. We choose $\tilde{g}$ so that \eqref{bfield10d} is canonically normalized, and hence $\tilde{g}^2V_6/(4\kappa_{10}^2)=\frac12$, where $V_6$ is the volume of the compactification manifold. With this choice of $\tilde{g}$, we may read the canonical axion decay constant $f$ from \eqref{axdecons}, to have 
\begin{align}\tilde{g}^2=\frac{2\kappa_{10}^2}{V_6} ,\quad f=\frac{2\pi\alpha'}{\tilde{g} V_{\Sigma_2}}=\sqrt{\frac{V_6}{2\kappa_{10}^2}}\frac{2\pi\alpha'}{V_{\Sigma_2}}.\end{align}

One may also get the value of $\Lambda_k^2$, the 4-form quantum, using the above and the DBI action. If we momentarily take $F_2=0$, we may obtain $g$ from the DBI by substituting $B_2=\tilde{g} \phi \omega_2$ and expanding to second order in $\phi$, from which we get
\begin{align} S_{DBI}\approx \frac12\frac{\mu_5}{g_s} V_{\Sigma_2}\tilde{g}^2\phi^2,\quad\Rightarrow\quad g^2=\frac{\mu_5}{g_s} V_{\Sigma_2}\tilde{g}^2=\frac{2\sqrt{\pi}}{4\pi^2\alpha'}\frac{V_{\Sigma_2}}{V_6 g_s}\kappa_{10}.\end{align}
From this, we get $\Lambda_k^4=(2\pi g f)^2$ as
\begin{align}\Lambda_k^4=\frac{4\pi^2\alpha'\sqrt{\pi}}{g_sV_{\Sigma_2}\kappa_{10}}=\sim\frac{M_s^2}{g_s V_{\Sigma_2}}.\label{lambdakvalor}\end{align}
This same value of $\Lambda_k^2$ can be obtained in a different way. Allowed values of the flux of the D5 gauge field $F_2$ are $F_2=\frac{2\pi n}{V_{\Sigma_2}}\omega_2$ for integer $q$, so that $\int_{\Sigma_2} F_2$ is an integer multiple of $2\pi$ (this ensures that the lower D-brane charges induced by the D-brane Chern-Simons terms satisfy the right Dirac quantization condition). Plugging this back into the DBI action, we get an action of the form $\frac12\Lambda_k^2 n^2$, with $\Lambda_k$ given by \eqref{lambdakvalor}, as expected.

Admittedly, we have no control over $\mu^2$, which is why the model in this section falls short from a stringy realization of relaxation. One possible way out  is to locate the relevant branes into  warped throats which can exponentially  reduce the mass scales $M,\sigma, m$ through a warping factor \cite{Randall:1999ee,Giddings:2001yu,Klebanov:2000hb}. This is
in fact used in KKLT-like moduli fixing models \cite{Kachru:2003aw}, in which this exponential suppression is also important in order to fine-tune the cosmological constant.
It is however not obvious that this proposal does not turn the relaxion proposal into a Randall-Sundrum solution of the hierarchy problem, as the quantum corrections to the Higgs mass $\mu^2$ will also be warped. We speculate that this may not be the case if $\mu^2$ receives contributions from sources outside of the throat; however, we have no explicit implementation of this idea, which is beyond the scope of this paper.

The WGC membrane must already be there in string theory. Indeed, this is the case: the membrane is the object coupling to the 4d 3-form, which is the magnetic dual potential of the gauge field on the D5-brane worldvolume. Hence, the WGC membrane is a monopole on the D5-brane worldvolume gauge theory, realized as a D3-brane ending\footnote{One may wonder where the other boundary of the D3-brane is. Actually, the worldvolume $U(1)$ gauge field whose dual has KS coupling must have no Stuckelberg couplings \cite{BerasaluceGonzalez:2012zn}, and the latter condition implies that the $U(1)$ resides on a combination of D-branes with a homology relation, which thus defines a chain \cite{Camara:2011jg,Marchesano:2014bia}, on which the monopole D-brane actually wraps.} on the D5-brane \cite{Hanany:1996ie}.

An additional improvement to achieve a more realistic model would be to realize the full SM gauge group and matter,  and achieve full moduli fixed. This would require a global embedding, in particular with cancellation of RR tadpoles, and hence possibly requiring anti-D5-branes, thus complicating the construction.
Note  finally  that the assumption of a 
large negative Higgs mass-squared from quantum corrections is crucial to have the relaxation effect, since otherwise the potential would have been
positive definite and the Higgs would never reach a zero mass point.     It would be interesting to address these issues and other ingredientts in other
classes of string vacua.  

In summary, string theory has in principle many of the ingredients of the simplest relaxion models. However, the detailed realization should address the difficulties of obtaining a viable example, in particular producing the very peculiar structure of mass scales in relaxion models.

\section{Conclusions}\label{sec:conclusions}

The idea that the cosmological evolution plays a role in the origin of hierarchies  and other physical properties in Particle Physics
is tantalising. The  work in \cite{Graham:2015cka} studies the generation of the EW scale in terms of the cosmological evolution of an
axion field, the relaxion. The simplest implementations based on this idea have  however a number of  problems, including the
inconsistent  explicit breaking of the axion discrete gauge symmetry, as well as the stability of extremely large trans-Planckian 
excursions of the relaxion.  It was pointed out from the beginning that a monodromy structure of the axion could allow for large 
field excursions, following the pattern of  large field monodromy inflation in string theory.  However no explicit  relaxion  model has
been constructed implementing this class of symmetries.

In this paper we have described  how one can construct relaxion type models with a monodromy structure built in.  The simplest way 
to achieve this is  by coupling a Minkowski 3-form, with a quantized flux  $F_4$  coupling both to the relaxion and   the Higgs field. 
The gauge symmetries of the 3-form guarantee the stability of the axion potential under Planck suppressed  corrections, similar to
the Kaloper-Sorbo structure applied in  large field inflation.  This allows for large trans-Planckian excursions of the 
relaxion field.   In these constructions  the axion shift comes along with shifts
of the $F_4$ quanta, so that there is a branched structure of axion potentials. The shift symmetry is in this way consistent with
the presence a non-vanishing potential for the relaxion.  An interesting consequence of the monodromy version of relaxion 
is that  the mass parameter $g$ is related to the 4-form flux by $g=F_4/f$, so that one may understand the smallness of the 
$g$ parameter from a sort of see-saw structure. Thus, e.g., values of the 4-form flux $F_4\simeq (10^{-3}$eV$)^2$ and axion
decay constants $f\simeq 10^{10}$ GeV give rise to values $g\simeq 10^{-34}$, in the ballpark of simplest relaxion models.
The fact that this  value  for  $F_4^2$ is of order of the observed cosmological constant $\Lambda_{dark}\simeq (10^{-3}$eV$)^4$ is
intriguing.

We also describe how in explicit string constructions the main ingredients of the relaxion mechanism are present. There are
axions arising from the dimensional reduction of RR and NS antisymmetric tensors and these axions may have couplings to
Higgs doublets  with a structure similar to relaxion models. A toy model from the DBI dynamics of D5-branes in Type IIB
orientifolds is presented.  However  the natural scale for the 4-forms in string theory is the string scale
$F_4\simeq M_s^2$, whereas realistic relaxion models have rather  very low values, e.g. $F_4\simeq  (10^{-3}$eV$)^2$.
It would be interesting to see whether the presence of strong warping effects or other mechanisms 
could explain this difference in scales.

We have studied  the constraints on the relaxion parameters coming from membrane nucleation.
In particular 
there are membranes coupling to the 3-forms which induce changes on the 4-form quanta, while also changing the axion branch.
These jumps make the slow roll of the relaxion unstable,   so that  we must impose that the rate for these  processes  
is sufficiently  suppressed.
To check whether the nucleation of membranes is suppressed or not we need to know the membrane tension T.  However
we have argued that the Weak Gravity Conjecture provides for an upper bound for the tension of order $T\leq gfM_p$. With this input
we find that the transition rate from one relaxion branch to another is indeed suppressed if the cut-off scale is below 
a certain scale ($M\lesssim 10^{9}$ GeV in the simplest axion model) which is comparable to other limits. For a particular simple model in which the relaxion is also the QCD axion, we obtain the more stringent bound $M\lesssim 2000$ TeV. While this is less stringent than other constraints in the model, it provides an example of a sharp constraint to effective field theory parameters coming from quantum gravitational effects.

\section*{Acknowledgments}

We thank J. Brown, S. Franco, F. Marchesano, D. Regalado, A. Retolaza, G. Shiu and P. Soler for useful discussions. This work is partially supported by the grants  FPA2012-32828 from the MINECO, the ERC Advanced Grant SPLE under contract ERC-2012-ADG-20120216-320421 and the grant SEV-2012-0249 of the ``Centro de Excelencia Severo Ochoa" Programme. M.M. is supported by a ``La Caixa'' Ph.D scholarship. I.V. is supported by a grant from the Max Planck Society. The authors thank the organizers of the String-Pheno-Cosmo 2015 workshop, held at the Galileo Galilei Institute in Florence, which sparkled our interest in the topic of the paper.  M. M. thanks the Department of Physics of UW-Madison for hospitality during the later stages of this work. 
\newpage

\appendix
\section{Axion monodromy in the dual 2-form view}
\label{sec:Kaloper-Sorbo}

The  system described in section \ref{sec:axion-monodromy} admits an alternative description in which the scalar is dualized into a 2-form. Let us recall the duality in the absence of monodromy. Consider an axion field with action
\begin{align}
\frac{1}{2}\int d\phi\wedge*d\phi.
\label{actax}
\end{align}
As is well-known, the above theory is dual to that of a massless 2-form field $B$, as follows (see e.g. \cite{Dvali:2005an}). The path integral for the axion field is equivalent to the path integral of a closed 1-form $v$. This is described by the action
\begin{align}
S=\int \frac{1}{2}v\wedge*v+ B_2\wedge dv.
\end{align}
where $v$ is an unconstrained $1$-form. Integrating out $B$, it acts as a Lagrange multiplier imposing $dv=0$, and by then setting $v=d\phi$ we recover  the original action. If on the other hand we integrate out $v$ we obtain the action
\begin{align}
S= \frac{1}{2}\int dB_2\wedge*dB_2.
\end{align}

The Kaloper-Sorbo proposal allows us to extend this duality framework for the case of an axion with a potential. Consider the particular case of an scalar field with a mass term 
\beqa
\frac{1}{2}\int d\phi\wedge*d\phi - \frac12m^2\phi^2.
\label{actax2}
\eeqa
To establish the duality \cite{Dvali:2005an}, notice that the path integral for massive axion action
can be rewritten (at least in flat space) as 
\begin{align}
\frac{1}{2}\int d\phi\wedge*d\phi+g\phi F_4 -\frac12 F_4\wedge *F_4.
\end{align}
Repeating the argument above for the axion, we then get 
\begin{align}
S=\int \frac12 \left|dB_2-gC_3\right|^2-\frac12 F_4\wedge *F_4.
\label{1}
\end{align}
 This describes a massless three-form $C$ becoming massive by eating up a massless two-form $B$. Note the gauge invariance
 \beqa
 C_3\to C_3+d\Lambda_2\quad ,\quad B_2\to B_2+g\Lambda_2
 \eeqa
 
 The relationship \eq{qmf} also arises in the dual formulation. The operator 
\begin{align}\exp\left(2\pi i f \int_{\partial\Sigma}B_2-k\Lambda^2_k\int_{\Sigma} C_3\right)\end{align}
 is gauge-invariant both under large $B$-field gauge transformations and $C$ gauge transformations, provided that $k\Lambda_k^2=2\pi gf$. Physically the operator describes the contribution of an euclidean membrane instanton wrapping $\Sigma$ and ending in an worldsheet instanton wrapping $\partial\Sigma$. $k\Lambda^2_k$ is then the membrane charge, which by Dirac quantization \cite{Bousso:2000xa} can be argued to be an integer multiple $k$ of the fundamental 3-form coupling, $\Lambda^2_k$.

\section{Analysis of the bubbles}\label{sec:bubbles}

Here we study in detail several aspects of 4-form membrane and bubble physics which are important for the main text. Specifically, we derive the tunneling probability formulae such as \eqref{decayrate}, and analyze the nucleation and growth of the bubbles, emphasizing the cosmological effects which take over when the bubble radius is comparable to the de Sitter radius.

\subsection{Coleman-DeLuccia formulae}
 We are interested in computing the nucleation probability in the thin-wall approximation for a membrane of tension $T$ which transitions from a vacuum with $V_i$ to another with $V_f$, $V_f<V_i$ and both positive. This is $e^{-B}$, and for $B$, there is a
slightly complicated formula by Coleman and DeLuccia \cite{Coleman:1980aw},
\begin{align}B=2\pi^2Tr^3+\frac{12\pi^2}{\kappa^2}\left\{\frac{1}{V_f}\left[(1-\frac13\kappa V_f r^2)^{\frac32}-1\right]-\frac{1}{V_i}\left[(1-\frac13\kappa V_i r^2)^{\frac32}-1\right]\right\},\label{bounce}\end{align}
where $\kappa\equiv 8\pi G$, and $M_P=G^{-1/2}$. One is supposed to minimize $B$ with respect to $r$, and then the bubble nucleation rate is $\exp[-B(r_{min})]$. The expression simplifies for $V_i=0$ and it has been used recently in the literature of WGC constraints to large field inflation \cite{Brown:2015iha}.
 
 However, we are interested in a different case, since for us $V_f=V_i-2\pi fg^2\phi_i$ and we assume $V_f-V_i$ small. Note that both $V_f$ and $V_i$ include contributions from the inflationary potential, and are indeed dominated by them. As a result, we will take
 \begin{align}\frac{\kappa}{3} V_f\approx \frac{\kappa}{3} V_i\approx H^2\end{align}
 in terms of the Hubble constant during inflation.
 
 Let us now find the extrema of $B(r)$. The condition for stationary point is, discarding the trivial one at $r=0$, 
 \begin{align}- \sqrt{1-r^2 \Lambda _f}+ \sqrt{1-r^2 \Lambda _i}+\frac{\kappa T}{2} r=0\end{align}
 where following Coleman's notation, $\Lambda_i=(\kappa V_i)/3$ and analogously $\Lambda_f$ are the corresponding cosmological constants.
 
 We now solve this equation carefully. Squaring once, we get
 \begin{align}\gamma r=-\kappa T  \sqrt{1-r^2 \Lambda _i},\quad \gamma\equiv \left(\frac{(\kappa T)^2}{4}+\Lambda_f-\Lambda_i\right).\label{sasas}\end{align}
 Existence of nontrivial solutions now depends crucially on $\gamma<0$. If this is not the case, the would-be bubble would have a radius so large it would extend beyond the cosmological horizon. Fortunately, in relaxation we have $\Lambda_f-\Lambda_i=-\frac{\kappa}{3}2\pi f g^2\phi_0$ with $\phi_0=M^2/g$. For this to overcome the WGC membrane tension $(\kappa T)^2\sim (fg/M_P)^2$, we must have $\phi_0>f$, the original vev of the relaxion must be larger than the fundamental decay constant, which is always satisfied in any reasonable relaxionic model.
 
One may wonder what happens for more generic monodromy models. For instance, \cite{McAllister:2008hb} provide examples of monodromic models which have a linear potential at large fields. For a generic potential, the requirement that bubbles can form is that the change $\Delta V$ in the potential upon crossing of a membrane $\Delta V\gtrsim (fg)^2$. Parametrizing a linear potential as $X^2g\phi$, for some mass scale $X$, we get a condition $\phi_0>f\frac{M}{X}$, which is only troublesome if $X\ll M$. However, both for the original relaxation proposal and the monodromic models in this paper, we take $X=M$. To do otherwise means introducing a new mass scale, $X$, very different from $M$. In a sense, this shifts the problem from explaining the hierarchy between $m_W$ and $M$ to explaining the hierarchy between $X$ and $M$.

 We now may take squares in \eq{sasas} to obtain
 \begin{align}r_{min}^2=\frac{1}{\left(\frac{\gamma}{\kappa T}\right)^2+\Lambda_i}\label{rsquared}\end{align}
 Plugging back in \eqref{bounce} we get our exact expression
 \begin{align}B=2 \pi ^2 \left(\frac{2 \left(\left(\frac{1}{\frac{\kappa ^2 T^2 \Lambda _f}{\tilde{\gamma }^2}+1}\right){}^{3/2}-1\right)}{\kappa  \Lambda _f}+\frac{T}{\left(\Lambda _i+\frac{\gamma ^2}{\kappa ^2 T^2}\right){}^{3/2}}-\frac{2 \left(\left(\frac{1}{\frac{\kappa ^2 T^2 \Lambda _i}{\gamma ^2}+1}\right){}^{3/2}-1\right)}{\kappa  \Lambda _i}\right)\label{B2}\end{align}
 where we have defined  
 \begin{align}\tilde{\gamma}\equiv \left(\frac{(\kappa T)^2}{4}+\Lambda_i-\Lambda_f\right).\end{align}
 When $\Lambda_i\rightarrow0$, it reduces to the transition-to-Minkowski limit
 \begin{align}B\rightarrow\frac{216 \pi T^4}{V (4 V + 3 \kappa^2T^2 )^2}.\end{align}
 We are instead interested in a different limit in which the difference $\Delta\Lambda=\Lambda_i-\Lambda_f$ is very small compared with either $\Lambda_i,\Lambda_f$, while also being very large compared to $(\kappa T)^2$. The parameter
 \begin{align}p\equiv\frac{\kappa T}{\sqrt{\Lambda}}=\frac{fg}{M_P H}\lesssim \frac{fg}{M^2}=\frac{f}{\phi_0}\end{align} 
 (where we have used $H\gtrsim M^2/M_p$, and renamed $\Lambda_i\equiv\Lambda$ as our reference cosmological constant) is very small in relaxionic models, so it is a nice variable to expand $B$. We also introduce
 \begin{align}q\equiv\frac{1}{\sqrt{\Lambda}}\left(\frac{\Delta\Lambda}{\kappa T}\right)\sim\frac{g\phi}{M_P H}\lesssim\frac{\phi}{\phi_0}.\label{defq}\end{align}
 Unlike $p$, $q$ can be of order $1$ or larger, but its introduction simplifies formulae\footnote{There is a numeric factor of $1/3$ in \eq{defq}, which we have ommitted. This does not affect our results, which are all order-of-magnitude estimates.}. $p$ is precisely the inverse of the parameter $A$ in \cite{Brown:2015iha}.

 After expansion for small $p$, we get
 \begin{align} 
 \kappa B=w(q)\frac{2 \pi ^2 p}{  \Lambda}+\mathcal{O}(p^2),\quad w(q)\equiv \frac{1+2q^2}{\sqrt{1+q^2}}-2q.
\end{align}
 The function $w(q)$ goes from $1$ at $q=0$ to $\approx 0.1$ at $q=1$. Substituting $p$, when studying membranes in relaxion scenarios, it is a good approximation to take
 \begin{align}B\approx\frac{2 \pi ^2 T}{H^3} w(q)
\end{align}
 which is precisely \eqref{decayrate}.
 Notice that the radius of the bubble is always large. From \eqref{rsquared},
 \begin{align}r_{min}^2=\frac{1}{\Lambda}\frac{1}{1+\left(\frac{p}{4}-q\right)^2},\end{align} 
 which is very close to the de Sitter radius $(\Lambda)^{-1/2}$ for small $p$ and $p/q$. 
 
 Substitution of typical relaxionic values, while also using $fgM^2\sim\Lambda_v^4$, $M_PH\gtrsim M^2$, yields
 \begin{align}B\lesssim 2\pi^2 w(q)\left(\frac{\Lambda_v^4M_p^4}{m_W^8}\right)\left(\frac{\Lambda_v}{\Lambda_{QCD}}\right)^4\left(\frac{m_W}{M}\right)^{8}\sim 10^{57} w(q)\left(\frac{\Lambda_v}{\Lambda_{QCD}}\right)^4\left(\frac{m_W}{M}\right)^{8}.\end{align}
 where we have used $\Lambda_{QCD}\sim0.2\ GeV$, $M_P\sim 10^{19}\ GeV$, $m_W\sim 10^2\ GeV$. This is another form of equation \eqref{prob2}, which highlights the enormous scales involved.
 
 A final but important comment is in order. The results of \cite{Coleman:1980aw} which we used as starting point rely on the thin-wall approximation. As discussed in that reference, this means that the thickness of the membrane $L$ should be much lower than the de Sitter radius $\Lambda^{-1/2}=H^{-1}$. Here we are at a loss: although the WGC gives the value of the tension of the membrane, it does not provide a value for $L$. 
 
We have no proof that the bubbles we consider satisfy the thin-wall approximation. However, we can give a plausibility argument.  Generically, we expect the effective field theory to be a valid description of the physics up to scales of order the cutoff $M^2$. If the membrane is thicker than $M^{-2}$, it should arise as a soliton of the effective field theory. However, there are no such solitons \footnote{Barring the field-theoretic bubbles which arise when the wiggles in the relaxion potential become large; as discussed in the main text, these are not the bubbles we are concerned with.}. So either the EFT we were using is incomplete, or we can trust the thin-wall approximation.

For the stringy models similar to the one in section \ref{sec:string-relaxion} we can be a little more explicit, since the membrane is generically a $D$-brane wrapping some cycle or chain in the compactification manifold. Then the membrane thickness will generically be of order $\sqrt{\alpha'}g_s$ \cite{Shenker:1995xq,Douglas:1996yp}. In any successful stringy embedding of relaxation, this must be much smaller than the de Sitter radius to claim control of the theory (otherwise the tower of excited string states and winding modes stretching around de Sitter space become light during relaxation). Otherwise we would have e.g. low tension fundamental strings which should be included in the effective field theory during inflation.
 
 In other words, if the thin-wall approximation for the WGC membrane does not hold, we likely cannot trust field theory for scales of order $\sim H$ anyway. Hence we expect to be able to trust the formulae of this section in the relaxionic context, if the relaxion proposal indeed admits an UV completion.   
 \subsection{Bubble growth, energy balance, and cosmological effects}
The bubble solution is obtained in the so-called hyperbolic coordinates \cite{Spradlin:2001pw}. One may change to global coordinates, more natural in the inflationary context, by recalling the definition of de Sitter space as the hyperboloid
\begin{align}\Lambda^{-2}=w^2-\tau^2+x^2+y^2+z^2=w^2-\tau^2+\rho^2\end{align}
as done in \cite{Coleman:1980aw}. These coordinates are related to the usual ones by
\begin{align}\Lambda \rho^2=-\sinh^2(\sqrt{\Lambda}t)+\cosh^2(\sqrt{\Lambda}t)\Lambda r^2\end{align}
where $r$ is a dimensionful coordinate such that the metric on the 3-sphere of $t=0$ is just the flat metric in $\mathbb{R}^4$ restricted to the locus $\Lambda^{-2}=w^2+r^2$. The worldvolume of the membrane boundary is given by $\rho=r_{min}$, where $r_{min}$ is given by \eqref{rsquared}. Therefore, the bubble is indeed a ball in global coordinates, whose radius expands according to
\begin{align}r(t)=\sqrt{\frac{r_{min}^2}{\cosh^2(\sqrt{\Lambda}t)}+ \frac{1}{\Lambda}\tanh^2(\sqrt{\Lambda}t)}.\end{align}
For small enough $r_{min}$, we recover the Minkowski limit, but for late times the membranes reach the cosmological horizon. At this point they are frozen by the exponential expansion of de Sitter space. Similarly, membranes initially larger than the de Sitter radius contract instead of expanding. This may give some insight as to why no bubbles exist for $\gamma>0$; for $\gamma=0$, $r_{min}$ is precisely $(\Lambda)^{-1/2}$, the de Sitter radius. Bubbles with $\gamma>0$ should therefore have an initial radius larger than this, and therefore would initially contract. But the critical bubble is precisely the smallest one which can expand; hence no critical bubbles exist for $\gamma>0$.

The expression \eqref{rsquared} tells us that gravitational effects make  the critical bubble radius smaller than it would have been otherwise: The flat-space expression
\begin{align}r^2_{\text{min,flat space}}=\left(\frac{ T}{\Delta \Lambda}\right)^2>\frac{1}{\left(\frac{\gamma}{\kappa T}\right)^2+\Lambda_i}= r^2_{min}. \label{rminflat}\end{align}
The flat-space expression can be obtained straightforwardly by demanding that the total energy of the bubble vanishes, since in tunneling (or in any other transition) energy is conserved. In the thin-wall approximation one can take
\begin{align}E\approx 4\pi T r^2-\frac43\pi r^3=0,\end{align}
which results in \eqref{rminflat}.

A similar argument holds when gravitational effects are taken into account. This is because in asymptotically de Sitter spacetimes the (ADM) mass is well-defined, thanks to the presence of a timelike Killing vector (within the horizon). Consider a spherical region of radius $r$ at a constant time slice in de Sitter. Initially, we have empty de Sitter space with cosmological constant $\Lambda_i$. The ADM mass of this region is given by \cite{Wald}
\begin{align}GM=\int_0^r 4\pi r^2\rho(r')dr'=\frac12 \Lambda_i r^3.\label{Minitial}\end{align}
This expression already takes into account the gravitational self-energy of the system (which precisely cancels the effect of the warping of the geometry within the bubble). Now, this region is replaced by a thin-wall bubble whose cosmological constant is $\Lambda_i$. We have a contribution $\frac{4}{3}\pi \Lambda_i r^3$ to the energy, analogously to \eqref{Minitial}. However, this is not the end of the story. The bubble is bounded by a membrane, which self-gravitates. If the tension of the membrane is $T$, close to it the energy density is $T_{00}=T\delta (l)$, where $l$ is the normalized normal coordinate to the membrane. We have $\delta(l)=\delta(r)(g_{rr})^{1/2}$. Since $(g_{rr})^{1/2}$ jumps precisely at the location of the membrane, its value can be taken as the arithmetic mean of the values at both sides of the membrane\footnote{There is some ambiguity here, as the product $\delta(r)(g_{rr})^{1/2}$ involves a product of two distributions which must be conveniently regularized. The arithmetic prescription used here comes from using exactly the same smoothing for the membrane $\delta$ as the one used for the change in vacuum energy.} . As a result, the total ADM mass of the bubble is
\begin{align}GM=\frac12\Lambda_f r^3+  \frac{\kappa T}{4} r^2\left(\sqrt{1-\Lambda_f r^2}+\sqrt{1-\Lambda_i r^2}\right).\label{Mfinal}\end{align}
Demanding equality with \eqref{Minitial} yields precisely \eqref{rsquared}. Since \eqref{Mfinal} is a full expression for the energy of the membrane including gravitational backreaction, we may expand it for small $\Lambda_i,\Lambda_f$ to see the effect of the first corrections:
\begin{align}GM\approx \frac{4}{3}\pi \Lambda_f r^3+4\pi^2Tr^2 -2\pi T\left(\frac{\Lambda_i+\Lambda_f}{2}\right)r^4 +\mathcal{O}(\Lambda^2).\end{align}
The $r^4$ term is precisely the Newtonian gravitational interaction energy between a membrane of tension $T$ and radius $r$ with a ball of energy density $\frac12(V_i+V_f)$. We now have a clear picture of what is happening: The membrane feels the gravitational field of the bubble interior, which pushes it inwards\footnote{We are neglecting the gravitational field sourced by the membrane itself, but it would be easy to include. All we would have to do is to replace the $\sqrt{1-\Lambda_i r^2}$ term above by $\sqrt{1-\Lambda_f r^2-2m/r}$, where $m=4\pi Tr^2$ is the ADM mass of the membrane alone. Nevertheless, in relaxion models with WGC membranes, this contribution is negligible, again of order $f/\phi_0$.}. The gravitational field sourced by the energy density outside of the bubble is zero via Gauss' law. For large bubbles, this gravitational attraction is strong enough so as to diminish the radius of the critical bubble significantly.

\section{Weak Gravity Conjecture for $(d-1)$-form fields}
\label{WGC3}
Important subtleties arise when trying to apply the WGC to $p=d-1$, where the corresponding objects are domain walls, as noted by \cite{Heidenreich:2015nta}. Let us particularize to $p=3$, $d=4$, for which the relevant object is a membrane in 4d, separating vacua with different value of 4-form flux, $F_4\rightarrow F_4+\Lambda_k^2$, and (since a three-form field is nondynamical \cite{Brown:1988kg,Duncan:1989ug,Feng:2000if,Bousso:2000xa,Dvali:2005an,Bielleman:2015ina}) thus different cosmological constant. In General Relativity, a solution describing a black membrane with nonvanishing cosmological constant $\Lambda$ is \cite{GP}
\begin{align}
ds^2=-Hdt^2+H^{-1}dr^2+r^2(dx^2+dy^2),\quad H\equiv -\frac{2m_T}{r}-\frac13\Lambda r^2.\end{align}
where $m_T$ is a measure of the brane tension, which should be taken positive.
Since $r$ is always timelike, this solution describes a time dependent background which is not asymptotically flat. Furthermore, the solution only has a single coordinate horizon,  at $r^3=-\frac{6m_T}{\Lambda}$, so for  $\Lambda>0$, the only coordinate horizon is  behind the singularity at $r=0$. In a sense, the membrane is always superextremal and hence the validity of its effective field theory description is questionable.

This fits with the naive extrapolation of the extremality condition for dilatonic membranes obtained in \cite{Heidenreich:2015nta}. For $p=d-1$ the extremality condition is given by
\begin{align}\left[\frac{\alpha^2}{2}-\frac{d-1}{d-2}\right]T^2\leq \frac{g^2 q^2 }{G_N}.\label{hrrest}\end{align}
Here $\alpha$ measures the strength of the dilaton coupling. We see that for $\alpha=0$ the equation has no solution; the membranes are always superextremal.

From \eq{hrrest}, it seems that a non-trivial dilaton profile can solve the pathologies of the membrane solution (as is the case for domain walls solutions in string theory). Alternatively, consistency of the membrane solution is sensitive to the particular matter coupled to gravity in effective field theory. This sensitivity prevents us from constructing a generic black membrane and studying its decay to formulate a WGC; for some theories it might well be that no nonpathological membranes can be constructed at all.

In string theory, there is another way to go around these issues and derive a WGC for membranes: one may use a chain of dualities, such as $T$ duality, which change the codimension of objects charged under $p$-form fields. This approach was pioneered in \cite{Brown:2015iha}, which used a chain of dualities to derive a WGC for $0$-form fields (producing bounds on instanton actions), which also escapes the naive extrapolation of the $p$-form WGC.

This will also work for $(d-1)$-forms fields coming from RR fields in string theory. In this case, the membranes will be $D$-branes wrapping some cycle of the compactification manifold. If we perform a $T$-duality along some circle direction, the $(d-1)$ form will become a $(d-2)$ form. Charged objects in this case are cosmic strings which, like the domain walls we were considering, cannot have flat asymptotics either. If we perform another $T$-duality along another circle direction, the $(d-2)$ form will turn to a $(d-3)$-form. Charged objects are $(d-3)$ branes, for which an extremality condition and WGC bound can be defined without further issue. Now we can $T$-dualize back. In this way, the original WGC for $(d-3)$ forms can be rewritten in terms of T-dual to give a formulation of WGC for $(d-1)$-forms.

To sum up, although the effective field theory  arguments which support the WGC for $p<d-2$ cannot be applied straightforwardly to domain walls,  these difficulties seem due to the high codimension of the charged object preventing a smooth solution within effective field theory. As soon as one includes a (strong enough) dilaton coupling or allows for stringy dualities, $p=d-1$ does not seem to be very different from the other cases. In the main text we explore the consequences of the WGC for  4d 3-form fields, in the particular context of relaxion models.

\subsection{WGC variants}

The discussion in the main text corresponds to the so-called \emph{mild} form of the WGC  \cite{ArkaniHamed:2006dz}. Its mildness follows because it only constrains the the charge-to-mass ratio of the state (domain wall), but it can be satisified with any value of $Q$. This does not constrain the low-energy effective theory: the conjecture can be satisfied by a domain wall with very large charge and tension, so large that it is not relevant for inflationary or relaxion dynamics. This is the membrane version of the loophole discussed recently in the literature in the inflationary context \cite{Rudelius:2015xta,Montero:2015ofa,Brown:2015iha}.

In \cite{ArkaniHamed:2006dz}, two other two possible versions of the WGC were considered:
\begin{itemize}
\item The state of \emph{least} charge under the $p$-form already satisfies \eq{WGCmild}; we will call this the \emph{first strong form} of the WGC.
\item The \emph{lightest} state charged under the $p$-form field already satisfies \eq{WGCmild}; this is the \emph{second strong form} of the WGC. 
\end{itemize}
These forms are not directly related to black hole arguments, and their validity is under
debate. The original WGC paper suggested a counterexample for the first strong form, which has however been deactivated in \cite{Heidenreich:2015nta}, by the argument that the state is actually charged under several $U(1)$'s. This motivated the proposal of the so-called Lattice WGC, valid for several $U(1)$'s, and which in fact implies both the first and second strong forms above when applied to a single $U(1)$.

The Lattice WGC \cite{Heidenreich:2015nta} states that for every point in the charge lattice there is an object with charge $Q$ satisfying \eqref{WGCmild}. In the main text, we assume the validity of this latter conjecture because it seems to be the only strong form which can be extended to several $U(1)$'s and be consistent under dimensional reduction.

Therefore, we explore the implications of demanding the existence of a light membrane of unit charge whenever we have a 3-form in the theory. Regardless of the WGC, this assumption holds in any stringy model we can think, and pose serious trouble for the relaxion proposal.

We also note in passing that, whereas for other $p$-form fields there is also a magnetic WGC which sets an unusually low cutoff scale for the theory, there is no magnetic version of the WGC for 3-forms in 4d, or in general for $(d-1)$-forms in $d$ spacetime dimensions. Thus, in principle there is no bound on how low $g$ can be (though as $g\rightarrow 0$, the membranes predicted by the electric WGC become tensionless).

\section{DBI  D5-potential for the axion}\label{sec:DBI}

The effective action for the microscopic fields of a system of D5-branes in the 10d Einstein frame is given by the Dirac-Born-Infeld (DBI) + Chern-Simons (CS) actions
\begin{multline}
S=-\mu_5 g_s^{-1}\textrm{STr}\int d^6\xi \,  \sqrt{-\textrm{det}\left(P[E_{MN}+E_{Mi}(Q^{-1}-\delta)^{ij}E_{jN}]+2\pi\alpha' F_{MN}\right) det(Q^i_j)}\\
+\mu_5\textrm{STr}\int P[C_6+C_4\wedge \CF_2]
\end{multline} 
where
\beqa
E_{MN}=g_s^{1/2}G_{MN}-B_{MN}\quad ;\quad Q^i_j=\delta^i_j+i2\pi\alpha'[\Phi^i,\Phi^k]E_{kj}\\
\sigma=2\pi\alpha'\quad ;\quad \mu_5=(2\pi)^{5}\alpha'^{-3}g_s^{-1}\quad ;\quad \CF_2=2\pi\alpha' F_2-B_2
\eeqa

$P[\cdot]$ denotes the pullback of the 10d background onto the D5-brane worldvolume and `STr' is the symmetrised trace over gauge indices. The indices $M,N$ denote the directions extended by the D5-brane while $i,j$ denote the transverse directions. In the absence of NS and RR fluxes the Chern-Simons action plays no
role in the discussion.

We are interested in the scalar potential for the position moduli of the D5's, since the Higgs field will later appear as off-diagonal fluctuations of the adjoint field parametrizing the position of a stack of D5-branes. Therefore we will neglect all the terms involving the 4d gauge bosons and the Wilson lines.
We are going to assume for the moment no warpping, diagonal Minkowski and compact metric and no mixed Minkowski-internal tensors. We also consider vanishing 3-form $G_3$ fluxes but allow for an open string background given by the magnetic worldvolume field strength
\beq
F_2=q\omega_2 \ ,
\eeq
where $\omega_2$ is the orientifold-odd volume form of the 2-cycle $\Sigma_2$ wrapped by the D5-brane. Even if there is no B-field induced by $G_3$ on the brane, we can still have a coupling of the D5 position moduli to the axion coming from dimensionally reducing $B_2$ in the same 2-cycle
\beq
B_2=\phi\omega_2 \ ,
\eeq
as we will see in the following.

Neglecting derivative couplings, the determinant in the DBI action  can be factorised between Minkowski and the internal space as follows
\beq
\textrm{det}(P[E_{MN}] + \sigma F_{MN})=g_s^{2}\,\textrm{det}\left(\eta_{\mu\nu}+2\sigma^2\partial_\mu\Phi\partial_\nu\bar\Phi\right) 
\cdot\textrm{det}\left(g_{ab} 
 +  g_s^{-1/2}\CF_{ab}\right)
\eeq
where $\mu,\nu$ label the 4d non-compact directions and $a,b$ the internal D7-brane dimensions, and 
\beq
\CF_{ab}=\sigma F_{ab}-B_{ab} \ .
\eeq

Then, using the matrix identity
\beqa
\label{det}
\textrm{det}(1+\varepsilon M)& = &1 + \varepsilon\,\textrm{tr }M - \varepsilon^2\left[\frac{1}{2}\textrm{tr }M^2 - 
\frac12(\textrm{tr }M)^2\right] \\ \nonumber
&+& \varepsilon^3\left[\frac13\textrm{tr }M^3-\frac12(\textrm{tr }M)(\textrm{tr }M^2)+\frac16(\textrm{tr }M)^3\right]\\ \nonumber
& - &\varepsilon^4\left[\frac14\textrm{tr }M^4-\frac18(\textrm{tr }M^2)^2-\frac13(\textrm{tr }M)(\textrm{tr }M^3)\right.\\ \nonumber
& + & \left.\frac14(\textrm{tr }M)^2(\textrm{tr }M^2)+\frac{1}{24}(\textrm{tr }M)^4\right] 
\eeqa
we obtain on one hand that 
\beq
-\textrm{det}\left(\eta_{\mu\nu}+2\sigma^2\partial_\mu\Phi\partial_\nu\bar\Phi\right)\textrm{det}\left(g_{ab}+g_s^{-1/2}\CF_{ab}\right)=\left(1+2\sigma^2\partial_\mu\Phi\partial^\mu\bar\Phi\right)\left(1+\frac12g_s^{-1}\CF_{ab}\CF^{ab} \right) \ ,
\label{DBIext}
\eeq
where we have neglected terms with more than two derivatives in Minkowski. On the other hand we have that 
\beq
det(Q^m_n)=g_s^2\textrm{det}\left(\delta^m_{n}+i\sigma[\phi^m,\phi^p]\delta_{pn}\right)=g_s^2\left( 1+\frac12\sigma^2[\phi^m,\phi^n][\phi_n,\phi_m]+\dots\right)
\eeq
In a supersymmetric configuration with vanishing D-terms $[\Phi, \Phi^*]=0$ we get
\beq
Tr([\phi^m,\phi^n]^2)=[\phi^m,\phi^n][\phi_n,\phi_m]=-4|[\Phi^1,\Phi^2]|^2
\eeq
and the quartic terms  combine with the quadratic terms to complete a perfect square (see e.g. \cite{Ibanez:2014swa,Grimm:2008dq}) .
To do the computation more explicit  we took the transverse space to the branes to be $T^4$, and $\Phi_1,\Phi_2$ the two adjoints 
parametrizing the position in the torus.

Putting everything together we find that the relevant part of the DBI action is given by
\beq
S_{DBI}=-\mu_5 g_s{\rm STr}\int d^6\xi \sqrt{ \left(1+2\sigma^2\partial_\mu\Phi^i\partial^\mu\bar\Phi^i\right)\left(1+\frac12g_s^{-1}\CF_{ab}\CF^{ab} \right)\left( 1-4\sigma^2|[\Phi^1,\Phi^2]|^2\right)^2} \ ,
\eeq
as used in the main text.

\bibliographystyle{jhep}
\bibliography{refs-relaxion}

\end{document}